\RequirePackage{fix-cm}
\documentclass[smallextended]{svjour3}       
\smartqed  
\usepackage{graphicx}
\usepackage{mathptmx}      
\usepackage{latexsym}
\usepackage[latin1]{inputenc} 
\graphicspath{ {images/} }
\usepackage{caption}
\usepackage{inputenc}
\usepackage{multirow}

\usepackage[ruled,vlined]{algorithm2e}

\usepackage{graphicx}
\usepackage[font=scriptsize]{subcaption}
\begin{document}

\title{The Impact of Political Party/Candidate on the Election Results From A Sentiment Analysis Perspective Using \#AnambraDecides2017 Tweets
}


\author{Ikechukwu Onyenwe       \and
        Samuel Nwagbo			\and
        Njideka Mbeledogu		\and
        Ebele Onyedinma			\and
}


\institute{I. Onyenwe \at
              Department of Computer Science, Nnamdi Azikiwe University, Nigeria \\
              \email{ie.onyenwe@unizik.edu.ng}           
           \and
           S. Nwagbo \at
           Department of Political Science, Nnamdi Azikiwe University, Nigeria
           \and
           N. Mbeledogu, E. Onyedinma \at
           Department of Computer Science, Nnamdi Azikiwe University, Nigeria \\
}

\date{Received: date / Accepted: date}

\maketitle

\begin{abstract}
This work investigates empirically the \textit{impact of political party control} over its candidates or vice versa on winning an election using a Natural Language Processing (NLP) technique called Sentiment Analysis (SA). To do this, a set of 7430 tweets bearing or related to \#AnambraDecides2017 was streamed during the November 18, 2017 Anambra State gubernatorial election. These are Twitter discussions on the top 5 political parties and their candidates termed \textbf{political actors} in this paper. We conduct polarity and subjectivity sentiment analyses on all the tweets considering time as a useful dimension of SA. Furthermore, we use the \textit{word frequency} to find words most associated to the political actors in a given time. We find most talked about topics using a topic modeling algorithm and how the computed sentiments and most frequent words are related to the topics per political actor. Among other things, we deduced from the experimental results that even though a political party serves as a platform that sales the personality of a candidate, the acceptance of the candidate/party adds to the winning of an election. For example, we found the winner of the election \textit{Willie Obiano} benefiting from the values his party share among the people of the State. Associating his name with his party \textit{All Progressive Grand Alliance (APGA)} displays more positive sentiments and the Subjective Sentiment Analysis indicates that Twitter users mentioning \textit{APGA} are less emotionally subjective in their tweets than the other parties.


\noindent
\textit{Disclaimer: This paper is an academic research. It is not politcally motivated}.
\end{abstract}

\section{Introduction}
\label{intro}

A political party, according to \cite{ACE2012}, is an organized group of people who exercise their legal rights to identify with a set of similar political aims and opinions, and one that seeks to influence public policy by allowing its candidates elected to public office. Political parties gain control over the government by winning elections with candidates they officially sponsor or nominate for positions in government. They also coordinate political campaigns and mobilize voters. Political parties exist to win elections to influence public policy. This requires them to build coalitions across a wide range of voters who share similar preferences. Even though the presentation of candidates and the electoral campaign are the functions that are most visible to the electorate(s), political parties fulfill many other vital roles which could directly or indirectly influence the people that are registered to vote (the voters)\footnote{Lumen. https://courses.lumenlearning.com/americangovernment/chapter/introduction-9/}. The appearance of candidates in the electoral campaign is widely accepted to also influence the election outcomes \cite{HOEGG2011}

The popularity of social networking, micro-blogging and blogging websites have evolved to become a varied kind of way people express their thoughts and feelings, and at such a huge quantity of data is generated. For example, the nature of micro-blogging allows users to post realtime messages about their opinions on a variety of topics, discuss current issues, complain, and express sentiments (positive/negative) about things that influence their daily life. Recently, people are sounding off online like never before, like checking the reviews or ratings of movies or products before watching the movie in theatres or buying the products. In fact, manufacturing companies and politicians have started to use micro-blogs as a medium to get a sense of general sentiments about the way people view their products, views or personalities. Hence, this paper investigates if the sentiment analysis of political data can discover insights that show the influence political parties have on their candidates or vice versa which may lead to their winning or losing an election. This is certainly interesting as it guides political parties/candidates to know if people support their program or not.
 
Social media and other information and computer technologies have changed the dynamics of politics and political participation in Nigeria since 2011. Political actors likewise political parties now reach out to wider political space in millions across climes with their manifestoes and ideologies without embarking on distance tours. Social media has added colouration to technics of the political campaign even in the developing countries. A larger percentage of our population has access to and are also consciously connected to social media for ideas and news sharing, information and entertainment \cite{Bettina2009,nwagbosnc2016}. \cite{nwagbosnc2016} further maintained that social media grants many people the chance to participate actively in political discourses by adding their views to issues under discussion.

In this study, we try to understand empirically from Twitter discussions if political parties or their candidates could influence winning or losing an election. For this purpose, we use Twitter data collected on the election day of the Anambra State Gubernatorial Election held on November 18, 2017. To do this,  we use a Natural Language Processing (NLP) method called Sentiment Analysis (SA) to conduct data analysis experiments on the election Twitter data. The experiments involve the polarity sentiment analysis (PSA) and the subjectivity sentiment analysis (SSA) on all the tweets considering time as a useful dimension of SA.

Our purpose of PSA and SSA is to find attitudes of the people towards the political actors and to evaluate whether Twitter users were tweeting facts during the election or whether most of their messages were emotional subjective opinions based on a given time. Furthermore, using the \textit{word frequency} and a topic modeling algorithm, we find words most associated to the political actors and most talked about topics and how they are related to each other per political actor in a given time.

Thus, to analyze tweets in terms of polarity and subjectivity, we propose the following research question:

\textbf{Research Question 1}: \textit{How sentiment of the tweets for a particular candidate/party behaves across a given time frame to ascertain attitudes of the public towards the political actors?}

\textbf{Research Question 2}: \textit{How subjectivity scores for each candidate/party varied across time and which of the candidate/party whose mention alone got a high frequency score in more subjective tweets?} 

Time has been considered in the literature as a useful dimension for sentiment analysis \cite{Giachanou20162,nguyen2012predicting}. Tracking opinion over time is a powerful tool that can be used for sentiment prediction or to detect the possible reasons for a sentiment change. In particular, understanding topic and sentiment evolution during election allow the government, election observers or people to capture sentiment changes and act promptly. For example, understanding the sentiment change on a particular candidate during an election can reveal possible topic trends that can show people's attitude about the candidate.

To evaluate the stated research questions, we performed the following experimental analysis on the Twitter data we collected (see Table \ref{tab:dataset}):

\begin{itemize}
\item Polarity and subjectivity analyses considering a two-hourly time granularity attribute. Then, the averages of the two analyses scores are calculated. Every two hours generally means tracking topic changes eight times a day from \textit{06:00 to 23:59}.
\item Find the most talked about topics. In each topic, we investigate a political actor's name with the highest frequency of occurrence. Investigation includes 
\begin{enumerate}
\item which of the political actor whose mention alone in a tweet got a high frequency score in more polarity or subjectivity tweets.
\item How important are the words most associating to a political actor in a given topic?
\end{enumerate}

\end{itemize}

The experiments started with preprocessing of the tweets and performing initial investigations on them to discover the most common co-occurring words and the number of tweets per candidate and political parties. Furthermore, we group the tweets based on the names of interest (top five political parties and candidates) and perform sentiment analysis using Textblob's Naive Bayes Classifier (NBC)\footnote{https://textblob.readthedocs.io/en/dev/} and SENTIWORDNET \cite{Esuli2007} on the set of tweets in each group to determine the polarity of each tweet. For finding most talked about topics and most frequently associating words, we use Latent dirichlet allocation (LDA)\cite{blei2003} and \textit{word frequency} respectively.

\section{Related Work}
\label{relatedlit}
In the modern politics, Twitter has been in the forefront of political discourse, with politicians choosing it as their platform for disseminating information to their constituents. This has instigated parties and their candidates to an online presence which is usually dedicated to social media coordinators.

In this section, we present some previous works related to sentiment analysis of Twitter discussions on politics. Sentiment  analysis is a Natural Language Processing task, where the system has to test the sentiments of texts based on the training data, which obviously sounds like a machine learning problem. Starting from being a document level classification task \cite{turney2002thumbs,pang2004sentimental}, it has been handled at the sentence level \cite{hu2004mining,kim2004determining} and  more  recently used in the analysis of political texts, especially from the Twitter collection. \cite{tumasjan2010predicting} validate Twitter as a forum for political deliberation and validly mirror offline political sentiment based on the context of the German federal election. \cite{Heredia2011} explores the effectiveness of social media as a resource for both polling and predicting the election outcome.  \cite{wang2012system} analyze public sentiment toward presidential candidates in the 2012 U.S. election as expressed on Twitter. \cite{vilares2015megaphone} ranks political leaders, parties and personalities for popularity by analysing Spanish political tweets. \cite{conover2011political} analysis of political polarization on twitter demonstrates that the network of political retweets exhibits a highly segregated partisan structure, with extremely limited connectivity between left- and right-leaning users. \cite{razzaq2014prediction} experimental results validate social media content as an effective indicator for capturing political behaviours of different parties. In other words, positive, negative and neutral behaviour of the party followers as well as the party's campaign impact can be predicted. \cite{Dahal2019} research shows that social media websites can be used as a data source for mining public opinion on a variety of subjects and LDA was applied for topic modeling to infer the different topics of discussion. \cite{Boutet2013} researches on the usefulness of analyzing Twitter messages to identify both the characteristics of political parties and the political leaning of users. \cite{Makazhanov2014} reveals in their work that the political preference of users can be predicted from their Twitter behaviour towards political parties. Furthermore, \cite{Pak2010TwitterAA} has shown that Twitter, a microblogging platform, is valid for building a corpus for sentiment analysis and opinion mining. \cite{deFran2018} proposes a method to segment the Twitter users into groups such as popular, activists and observers to help filter out information and give a more detailed analysis of the important events.

These researches demonstrated that political insight is a phenomenon present on Twitter, hence, this paper presents a comprehensive sentiment analysis considering the common co-occurring tweet words and polarized tweets connections among such groups as a political party, candidate and political party cum its candidates.

\section{Methodology}
\begin{figure}[!htb]
\centering
\includegraphics[width=0.6\textwidth]{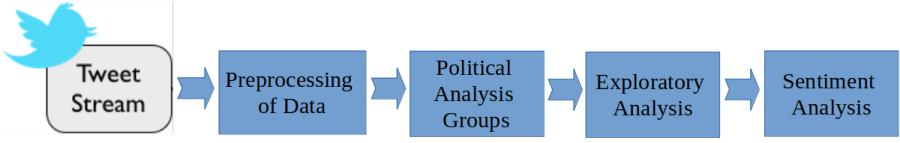}
\caption{\scriptsize Methodology steps}
\label{fig:methodsteps}
\end{figure}

Figure \ref{fig:methodsteps} shows the steps of the methodology followed in this work. In the first step, we collect Twitter data and described the process of tweets collection that formed the data of this work. We perform some clean up on the data collected as the second step. This process is discussed in detail in section \ref{preprocessing}. The third step is the political groups' analyses in section \ref{preprocessing} which are the groups of data recorded in the \textit{collection} column of Table \ref{tab:dataset}. The fourth and fifth steps of the methodology are analysed in \ref{explysis} and \ref{sentilysis} sections respectively. We perform the tweets' texts exploratory and sentiment analyses for three distinct groups: the individual parties, the candidates and the individual parties cum their candidates.

\subsection{Data Collection}
\label{datacll}
This section presents information about Anambra State and its gubernatorial election in the 18th November 2017, and the Twitter social network and its features. The method of collecting the November 18, 2017, Anambra State gubernatorial election Twitter data is discussed in the following subsections.

\subsubsection{November 18, 2017 Anambra State Gubernatorial Election}
\label{AGEactors}
Anambra is a state in southeastern Nigeria\footnote{https://en.wikipedia.org/wiki/Anambra\_State} with 21 Local Government Areas (LGAs). The State Gubernatorial Election (SGE) is conducted every 4 years just like every other state in the country. The November 18, 2017, SGE is a bit significant since the state became the first in the nation to have 37 political parties and candidates participated in the governorship election. In this paper, we only looked at the five major parties and their candidates: Willie Obiano (the incumbent Governor) of the state ruling All Progressive Grand Alliance (APGA), Tony Nwoye of the national ruling All Progressives Congress (APC), Oseloka Obaze of the People's Democratic Party (PDP), Osita Chidoka of United Progressive Party (UPP), and Godwin Ezeemo of the People's Progressive Alliance (PPA). The APGA candidate swept the entire 21 LGAs in the state according to the election results pulling a total vote of 234,071 to finish ahead of the candidate of APC who got 98,752\footnote{http://saharareporters.com/2017/11/19/governor-willie-obiano-wins-anambra-gubernatorial-election}$^,$\footnote{https://www.vanguardngr.com/2017/11/anambra-election-results-obiano-wins-21-lgas/}. This is of considerable significance in this research since the candidates of the other political parties involved are from some of the 21 LGAs.

\subsubsection{Twitter}
Twitter\footnote{http://twitter.com} is a social network classified as a microblog with which users can share messages, links to external websites, images, or videos that are visible to other users subscribed to the service. Messages that are posted on microblogs are short in contrast to traditional blogs. Blogging becomes 'micro' by shrinking it down to its bare essence and relaying the heart of the message and communicating the necessary as quickly as possible in realtime. Twitter, in 2016, limited its messages to 140 characters \cite{Giachanou2016}. There are other microblogging platforms such as Tumblr\footnote{https://www.tumblr.com/}, FourSquare\footnote{https://foursquare.com/}, Google+\footnote{http://plus.google.com.}, and LinkedIn\footnote{http://inkedin.com/} of which Twitter is the most popular microblog launched in 2006 and since then has attracted a large number of users. Researches, as presented in the \textit{Related Work} section, have shown that Twitter data is well suited as a corpus for sentiment analysis and opinion mining.

\subsubsection{Collecting data from Twitter}
\label{tweetcoll}
As discussed in Crawling Twitter Data of \cite{Shamanth2014}, data collection was done using Twitter Streaming Application Programming Interface (API) and Python. API is a tool that makes the interaction with computer programs and web services easy. It enables real-time collection of tweets. Many web services provide APIs to developers to interact with their services and to access data in a programmatic way. For this  work, we use the Twitter Streaming API to download tweets related to 3 keywords: ``\#anambradecides2017'', ``\#anambraelections'' and ``\#anambradecides'' on the day of the election. The objective of the real-time collection was to collect only tweets about the election published on the same day. We based on the hypothesis that \textit{if there is a tweet about Anambra State election that same day, then that tweet could be making a reference to what the user is experiencing at the moment about the election}. The Twitter data that we collected is stored in JSON format to make it easy for humans and computer to read from the data and to parse it respectively.

\subsection{Preprocessing Tweets Data}
\label{preprocessing}
\begin{figure}[!htb]
\centering
\includegraphics[width=0.6\textwidth]{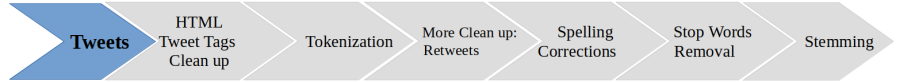}
\caption{\scriptsize Basic steps of preprocessing standard tweets}
\label{fig:preprocessingsteps}
\end{figure}

Figure \ref{fig:preprocessingsteps} shows five basic steps we took in preprocessing the dataset of tweets we collected as discussed in \ref{tweetcoll} section. 

\begin{table}[!htb]
\footnotesize
\caption{\scriptsize Dataset used in this study.}
\label{tab:dataset}
\centering
\begin{tabular}{l|l|r|r}
\hline
SN	&	Politcal Actors 		& 	Total of  						& Total of 				\\
	&	Tweets Collection		&	Tweets Before 					& Tweets After 			\\
	&							&	Preprocessing					& Preprocessing			\\
\hline
1	&	willie\_obiano 			&	1056							& 535					\\
2	&	oseloka\_obaze 			&	249								& 147					\\
3	&	nwoye\_tony 			&	602								& 235					\\
4	&	godwin\_ezeemo 			&	118								& 77					\\
5	&	osita\_chidoka 			&	277								& 118					\\
6	&	apga 					&	1728							& 668					\\
7	&	pdp 					&	1134							& 412					\\
8	&	apc 					&	1980							& 394					\\
9	&	ppa 					&	94								& 39					\\
10	&	upp 					&	2								& 2						\\
11	&	willie\_obiano\_apga 	&	2543							& 1079					\\
12	&	oseloka\_obaze\_pdp 	&	1258							& 489					\\
13	&	nwoye\_tony\_apc 		&	2464							& 577					\\
14	&	godwin\_ezeemo\_ppa 	&	160								& 92					\\
15	&	osita\_chidoka\_upp 	&	279								& 120					\\
\hline
16	&	Total Tweets			&	33502							& 7430					\\
\hline
\end{tabular}
\end{table}

We did preprocessing in two ways: \textit{Method 1} involves using tweet-preprocessor, a preprocessing library for a tweet data written in Python, to clean and tokenize the tweets. Tokenization involves converting a sentence into a list of words. In \textit{method 2}, we manually defined a function to double-check our tweet preprocessing and remove other unwanted tweets like retweets. This is to be sure that our data is reasonably cleaned. Moreover, spelling correction is one of the unique functionalities of the TextBlob library. With the correct method of the TextBlob object, we corrected all the spelling mistakes in our tweets. The final steps involve removing stopwords and punctuations, and stemming which is transforming any form of a word to its root word. Also included in the lists of stopwords are the party and candidate names, especially when we want to generate a word cloud image on any of the political parties or candidates since they are the targets. This is to enable meaningful words to be displayed than having party/candidate names seen all over the word cloud image.

Table \ref{tab:dataset} shows the total of tweets we collected and tweets associated with each political party and candidates before and after preprocessing. Rows 1 to 15 show the names we are interested in investigating on this study, we add columns of Booleans that indicated whether a name of interest was in the tweet or not. The \textit{Total of Tweets After Preprocessing} column shows that the names of interest for this work formed  67.08\% of 7430 total tweets. The remaining percentage is unclassified tweets. This experiment focuses on investigating whether the political actors stated in section \ref{AGEactors} can influence winning or losing an election. The results of the preprocessed tweets are stored in the CSV file. CSV file enables data storage into columns of variables and rows of observations.

\subsection{Experimental Tools}
\label{exptools}
In this experimental analysis, we use Sentiwordnet\cite{Esuli2007} to compare the overall analysis scores of TextBlob's Naive Bayes Classifier (NBC). The  two sentiment classifiers are used to determine the overall polarity scores for the sake of comparison, while Textblob is further used to perform detailed polarity analysis on the political actors and to determine the subjectivity of tweets. LDA (short for Latent Dirichlet Allocation) and  word frequency are used for topic modeling to infer the different topics of discussion and to find most common occurring words respectively. 

TextBlob is an extremely powerful NLP library for Python for processing textual data. It provides a consistent API for diving into common Natural Language Processing (NLP) tasks such as part-of-speech tagging, noun phrase extraction, sentiment analysis, and more. NBC is a classification technique based on Bayes'theorem with an assumption of independence among predictors. In simple terms, a NBC assumes that the presence of a particular feature in a class is unrelated to the presence of any other feature. NBC is based on the Bayes' theorem:  

\begin{center}
$P(A|B) = \frac{P(B|A)~*~P(A)}{P(B)}$
\end{center}

SENTIWORDNET \cite{Esuli2007} is the result of the automatic  annotation of all the synsets\footnote{A special kind of a simple interface that is present in NLTK to look up words in WordNet. } of WORDNET according to the notions of `positivity',  `negativity', and `neutrality'. Each synset \textit{s} is associated to three numerical scores \textit{Pos(s)}, \textit{ Neg(s)}, and \textit{Obj(s)} which indicate how  positive, negative, and ``neutral'' the terms contained in the synset are. Different senses of the same term  may thus have different opinion-related properties. For example, in SENTIWORDNET 1.0    the synset [estimable(J,3)] corresponding to the sense ``may be computed or estimated'' of the adjective estimable, has an  Obj score of 1:0 (and Pos and Neg scores of 0.0), while the synset [estimable(J,1)] corresponding to the sense ``deserving of respect or high regard'' has a Pos score of 0:75, a Neg score of 0:0, and an Obj score of 0:25. Each of the three scores  ranges in the interval [0:0;  1:0], and their sum is 1:0 for each synset. This means that a synset may have nonzero scores  for all the three categories, which would indicate that the corresponding terms have, in the sense indicated by the synset,  each of the three opinions related properties to a certain degree.

LDA \cite{blei2003} is an unsupervised machine-learning model that takes documents as input and finds topics as output. The model also says in what percentage each document talks about each topic. Hence, a topic is represented as a weighted list of words.

\subsection{Exploratory Twitter Data Analysis - EDA}
\label{explysis}
We use EDA approach to analyse the \textit{Total tweets} in Table \ref{tab:dataset} after preprocessing to summarize their main characteristics with visualizations. The EDA process is a necessary step prior to sentiment analysis or building a model in order to unravel various insights that will become important later. 

\subsection{Sentiment Analysis}
\label{sentilysis}

Sentiment analysis of tweets involves understanding the attitudes, opinions, views and emotions from tweets using Natural Language Processing (NLP) techniques. In this section, we look at sentiment analysis involving subjectivity and polarity. 

\subsubsection{Polarity and Subjectivity Analyses} 
Polarity is a sentiment analysis that determines whether a tweet expresses a positive, negative or neutral opinions. This enables the determination of the attitude of Twitter users for topics under discussion via quantifying the sentiment of texts.

Subjectivity is a sentiment analysis that classifies a text as opinionated or not opinionated. Terms such as adjectives, adverbs and some group of verbs and nouns are used to identify a subjective opinion. Speech patterns such as the use of adjectives along with nouns are used as an indicator for the subjectivity of a statement \cite{kharde2016sentiment,yaqub2018analysis}. Thus subjectivity analysis is the classification of sentences as subjective opinions or objective facts.

In this study, we have used tools as described in section \ref{exptools} on our dataset.  Hence from the dataset in Table \ref{tab:dataset}, we classify tweets as positive, negative or neutral opinions and further identify the ones that are subjective from those that are objective. Furthermore, we compute word frequency to find most talked about words, identify most discussed topics using LDA and look at how these most frequent words contribute to the importance of the LDA topics. Finally, we describe how the most frequent words and most discussed topics are related to the computed sentiments per political actor at a given time.

\begin{figure}[!htb]
\centering
\includegraphics[width=0.4\textwidth]{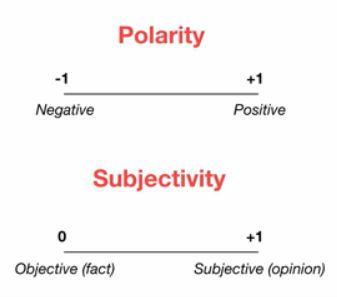}
\caption{\scriptsize Polarity and subjectivity metrics}
\label{fig:polmetrics}
\end{figure}

Figure \ref{fig:polmetrics} briefly describes the polarity metrics. Algorithm 1 shows subjectivity and polarity calculations. As explained in section \ref{exptools}, in Polarity, we are looking at how positive or negative a tweet is. -1 is very negative while +1 is very positive. For subjectivity, we are looking at how subjective or opinionated a tweet is. 0 is a fact while +1 is very much opinion.

\begin{algorithm}[H]
\small
\SetAlgoLined
 	polarity\_vals = []\;
	subjectivity\_vals = []\;
 \While{While lenght of tweets dataframe is not zero}{
  senti = SentimentClassifier(tweet)\;
   polarity\_vals.append(senti.sentiment.polarity)\;
   subjectivity\_vals.append(senti.sentiment.subjectivity)\;
  }
 \caption{ \scriptsize Algorithm to calculate the polarity and subjectivity of each tweet.}
\end{algorithm}

\section{Results}
This section presents the results of our experiments based on the stated research questions in the introduction.  While we used sentiment analysis to research on people's attitude towards political candidates and parties and whether such attitude is subjective or not, the use of exploratory Data Analysis gives us quantitative clues on our Twitter dataset.

\subsection{Exploratory Data Analysis (EDA)}

\begin{figure}[!htb]
\begin{subfigure}{0.5\textwidth}
  \centering
  \includegraphics[width=0.7\textwidth]{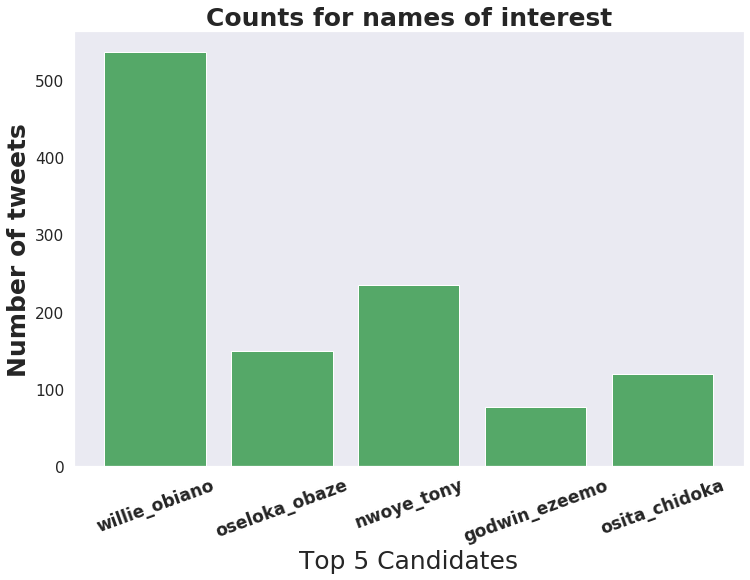}
  \caption{}
  \label{fig:namecounts1}
\end{subfigure}
\begin{subfigure}{0.5\textwidth}
  \centering
  \includegraphics[width=0.7\textwidth]{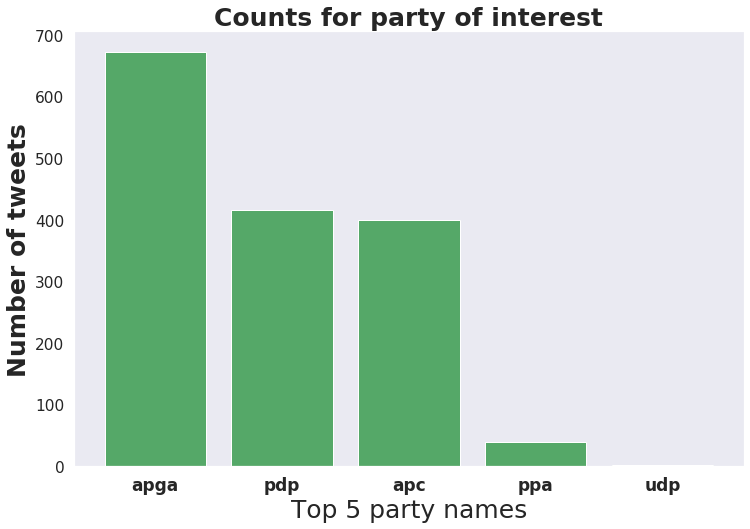}
  \caption{}
  \label{fig:namecounts2}
\end{subfigure}
\begin{subfigure}{0.5\textwidth}
  \centering
  \includegraphics[width=0.7\textwidth]{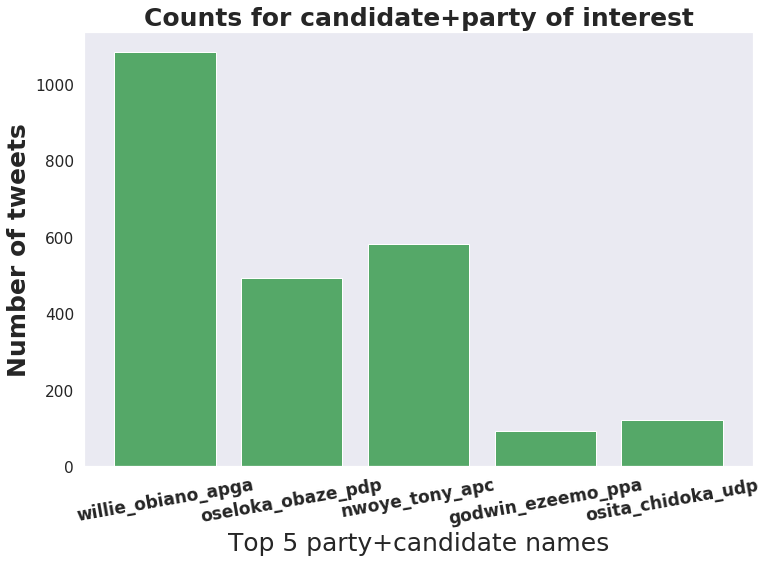}
  \caption{}
  \label{fig:namecounts3}
\end{subfigure}
\caption{\scriptsize A plot for counts for names of interest. (a) is counts for cadidates' names of interest. (b) is count for political parties' names of interest. (c) count for both parties' and candidates' names of interest}
\label{fig:namecounts}
\end{figure}

This process, as explained earlier in \ref{explysis}, is an important step usually performed before sentiment analysis to quantify the dataset in frequency. We start by counting names of interests by adding columns of Booleans in our Pandas data frame (a two-dimensional size-mutable, potentially heterogeneous tabular data structure with labelled axes (rows and columns)) to indicate whether a name of interest was in the tweet or not. See results in Table \ref{tab:dataset} from row 2 to 16 and Figure \ref{fig:namecounts}.

\begin{figure}[htb]
\begin{subfigure}{0.5\textwidth}
  \centering
  \includegraphics[width=0.6\textwidth]{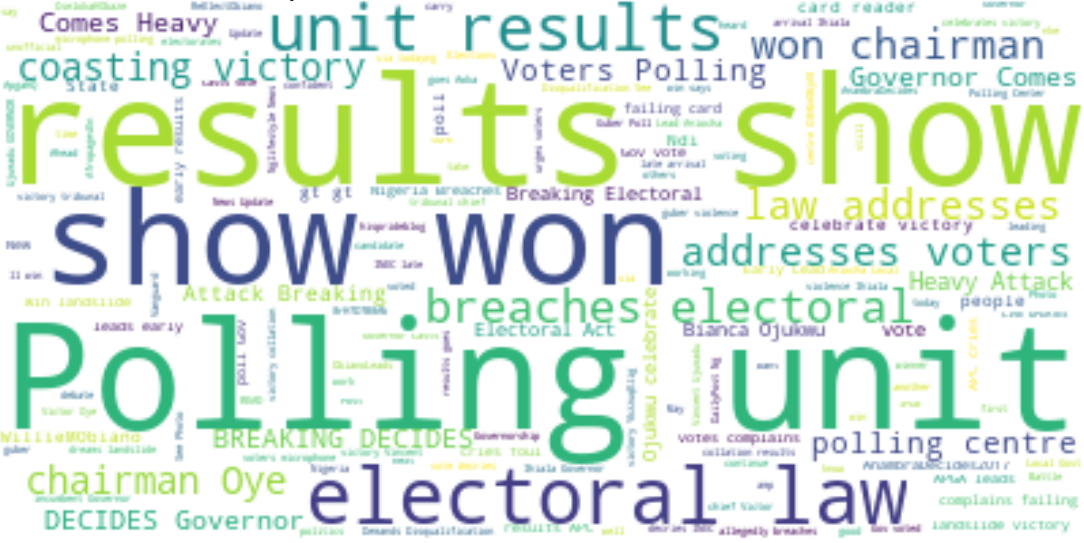}
  \caption{}
  \label{fig:wordcl_obiano}
\end{subfigure}
\begin{subfigure}{0.5\textwidth}
  \centering
  \includegraphics[width=0.6\textwidth]{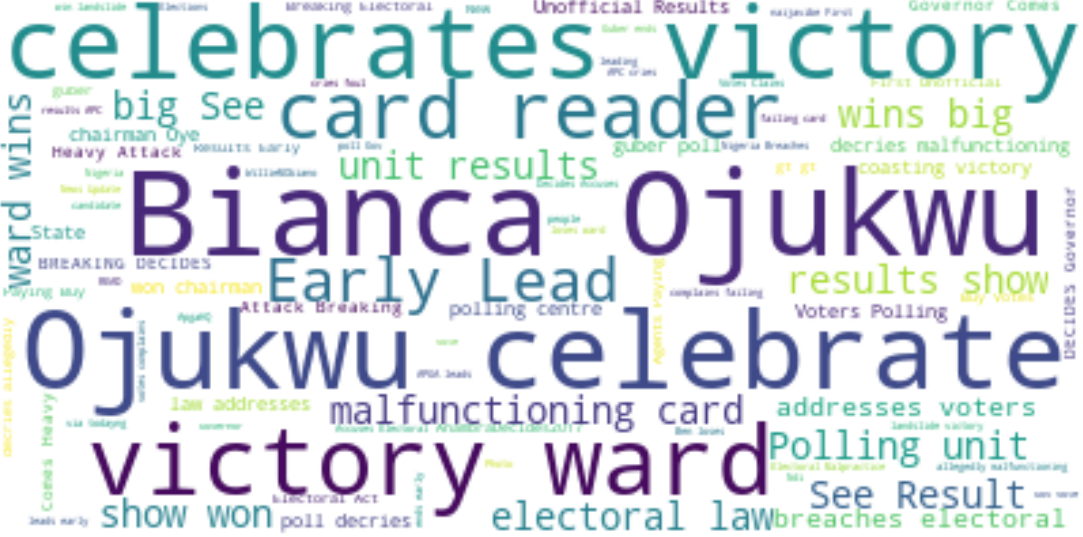}
  \caption{}
  \label{fig:wordcl_obianoapga}
\end{subfigure}
\begin{subfigure}{0.5\textwidth}
  \centering
  \includegraphics[width=0.6\textwidth]{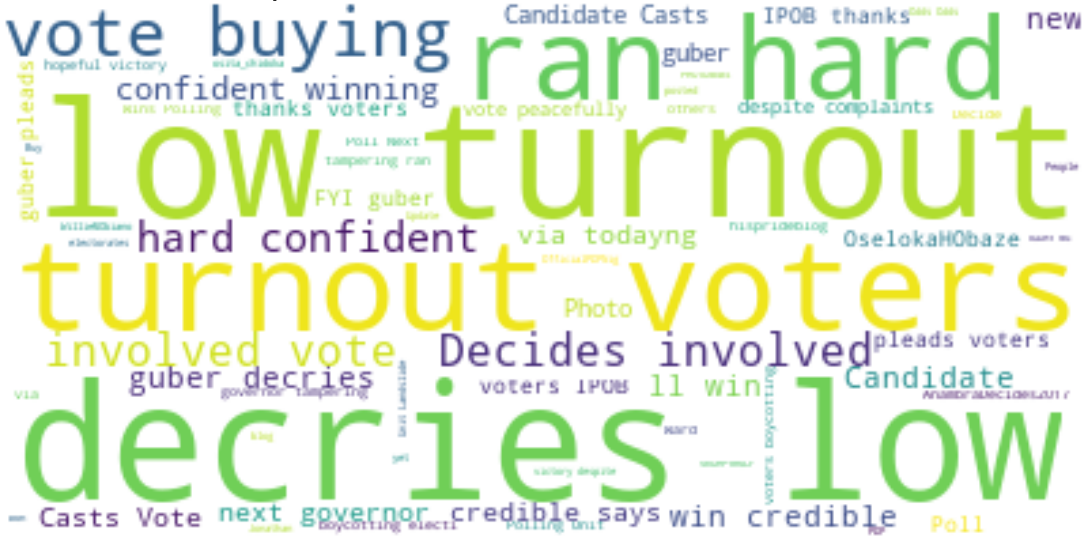}
  \caption{}
  \label{fig:wordcl_obaze}
\end{subfigure}
\begin{subfigure}{0.5\textwidth}
  \centering
  \includegraphics[width=0.6\textwidth]{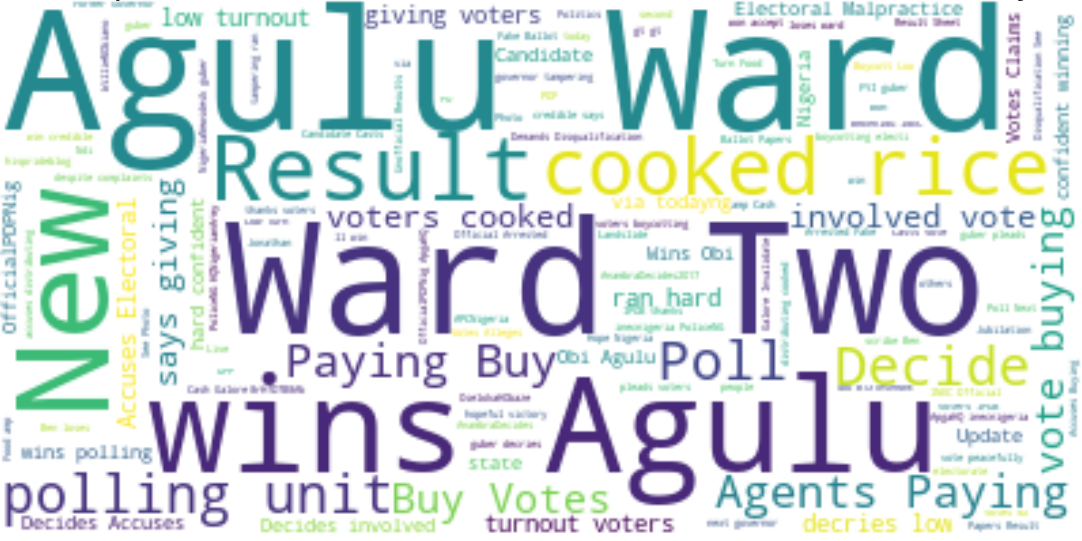}
  \caption{}
  \label{fig:wordcl_obazepdp}
\end{subfigure}
\begin{subfigure}{0.5\textwidth}
  \centering
  \includegraphics[width=0.5\textwidth]{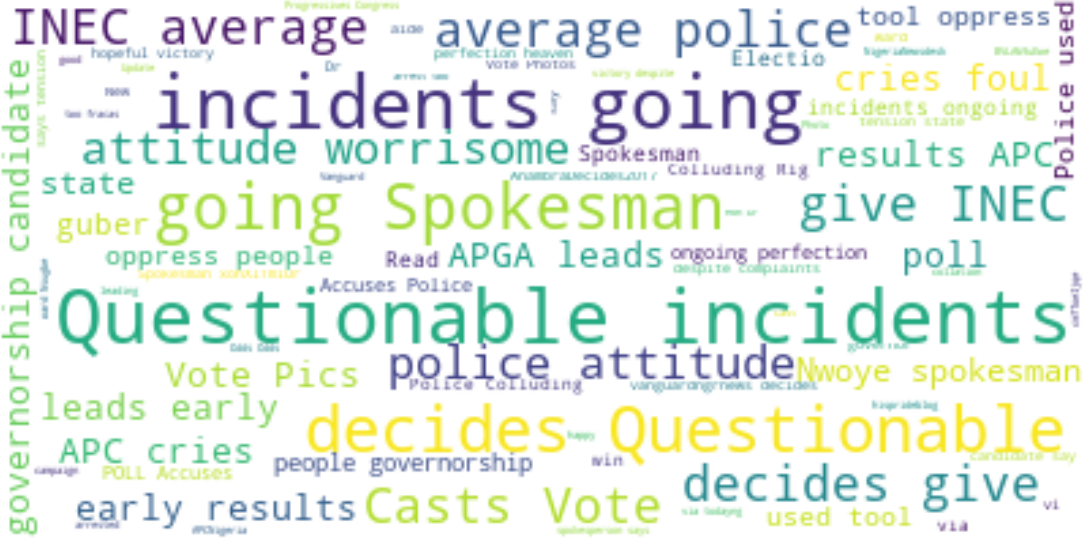}
  \caption{}
  \label{fig:wordcl_tony}
\end{subfigure}
\begin{subfigure}{0.5\textwidth}
  \centering
  \includegraphics[width=0.5\textwidth]{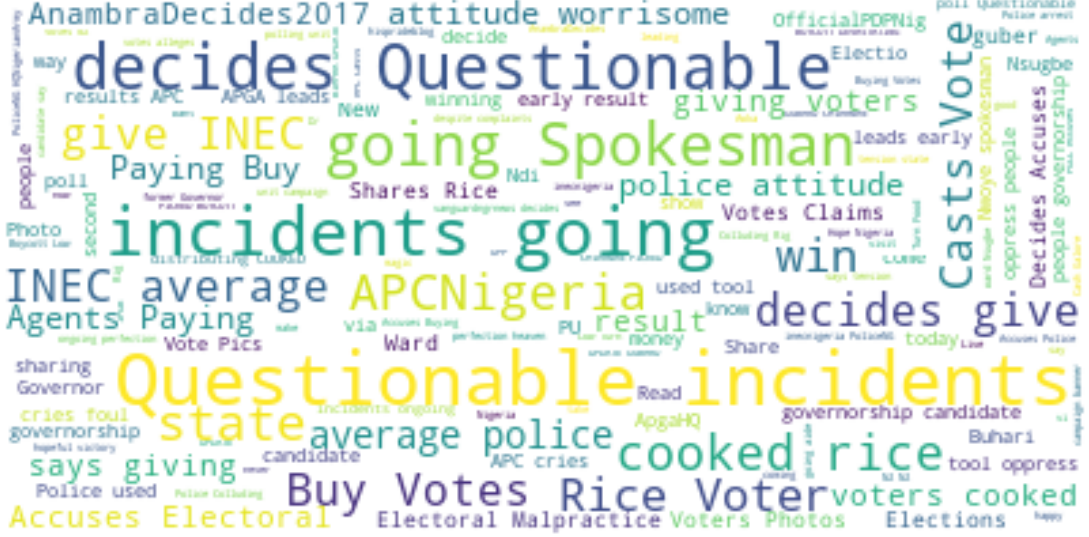}
  \caption{}
  \label{fig:wordcl_tonyapc}
\end{subfigure}
\begin{subfigure}{0.5\textwidth}
  \centering
  \includegraphics[width=0.5\textwidth]{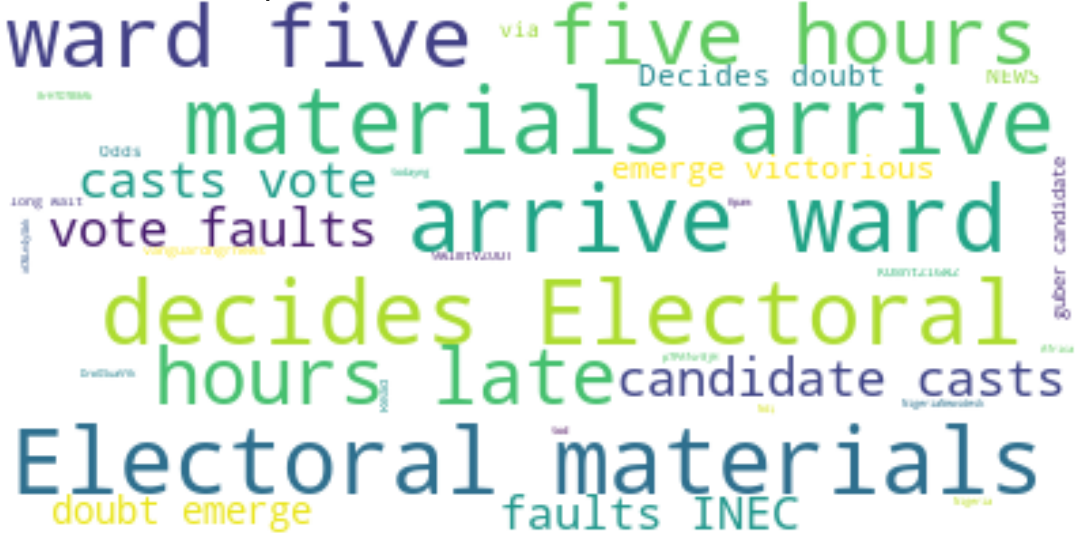}
  \caption{}
  \label{fig:wordcl_godwin}
\end{subfigure}
\begin{subfigure}{0.5\textwidth}
  \centering
  \includegraphics[width=0.5\textwidth]{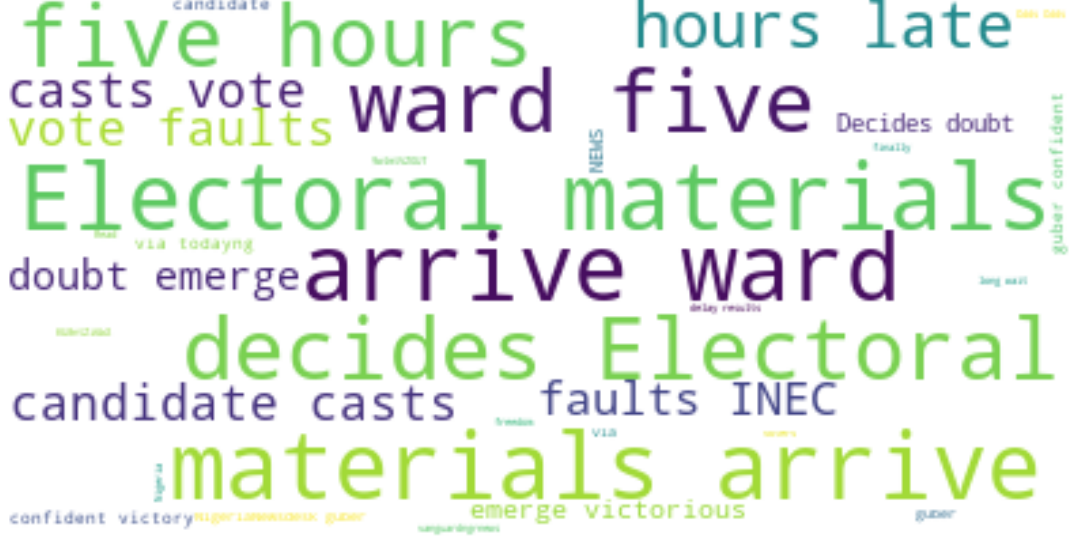}
  \caption{}
  \label{fig:wordcl_godwinppa}
\end{subfigure}
\begin{subfigure}{0.5\textwidth}
  \centering
  \includegraphics[width=0.5\textwidth]{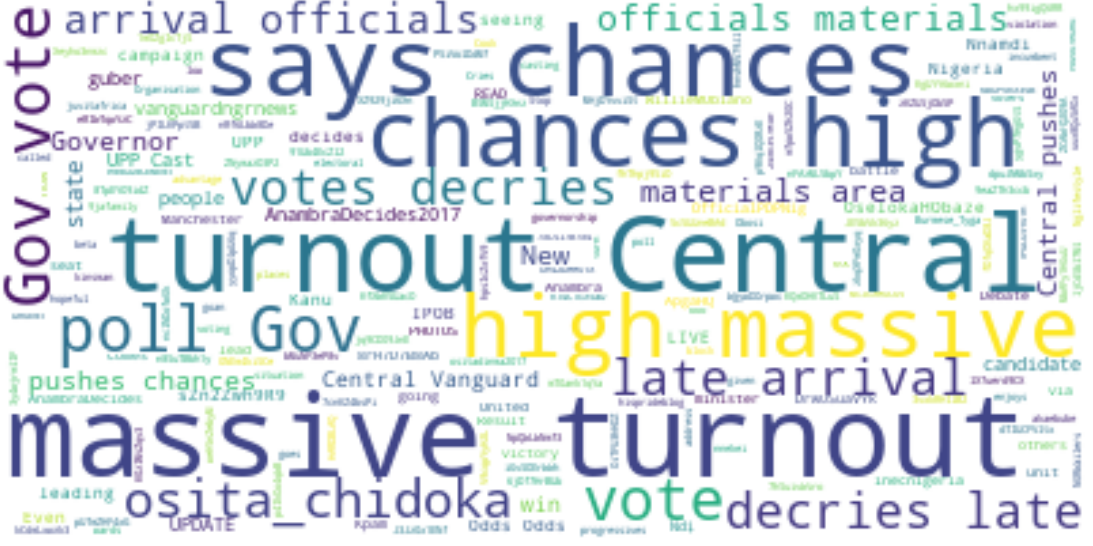}
  \caption{}
  \label{fig:wordcl_chidioka}
\end{subfigure}
\begin{subfigure}{0.5\textwidth}
  \centering
  \includegraphics[width=0.5\textwidth]{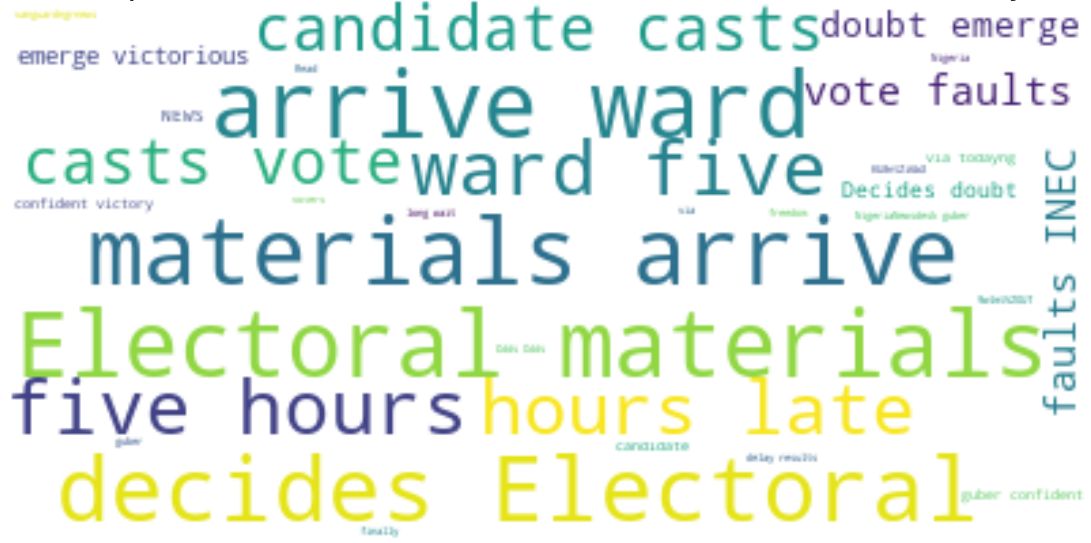}
  \caption{}
  \label{fig:wordcl_chidiokaudp}
\end{subfigure}
\caption{\scriptsize Words that co-occurred most frequently with the candidates and their parties. (a)-- Most frequent words associated with \textit{Willie Obiano} - the incumbent governor and winner of the election. (b)-- Most frequent words associated with \textit{Willie Obiano cum his party, APGA}. (c)-- Most frequent words associated with \textit{Obaze Oseloka}. (d)-- Most frequent words associated with \textit{Obaze Oseloka cum his party PDP}. (e)-- Most frequent words associated with \textit{Tony Nwoye}. (f)-- Most frequent words associated with \textit{Tony Nwoye cum his party APC}. (g)-- Most frequent words associated with \textit{Godwin Ezeemo}. (h)-- Most frequent words associated with \textit{Godwin Ezeemo cum his party PPA}. (i)-- Most frequent words associated with \textit{Osita Chidioka}. (j)--Most frequent words associated with \textit{Osita Chidioka cum his party UPP}.}
\label{fig:wordfreqCloud}
\end{figure}

Furthermore, we explore the most frequent words associated with the political actors. We added columns for tokens in our Pandas data frame and get rid of stopwords including punctuation, political parties and candidates names, then we generate a word cloud image for the frequent words. For example, words that co-occurred most frequently with the political actors\footnote{Political candidates and their parties} are shown in Figure \ref{fig:wordfreqCloud}.

\subsection{Sentiment Analysis}
At the first step of the sentiment analysis, we analyzed tweets using two different sentiment classifiers such as TextBlob and SentiWordNet with the aim of looking at their overall scores. The sole purpose of using these sentiment classifiers  was to give a comparison of their scores and to determine which classifier to use.

\begin{figure}[!htb]
\centering
\includegraphics[width=0.5\textwidth]{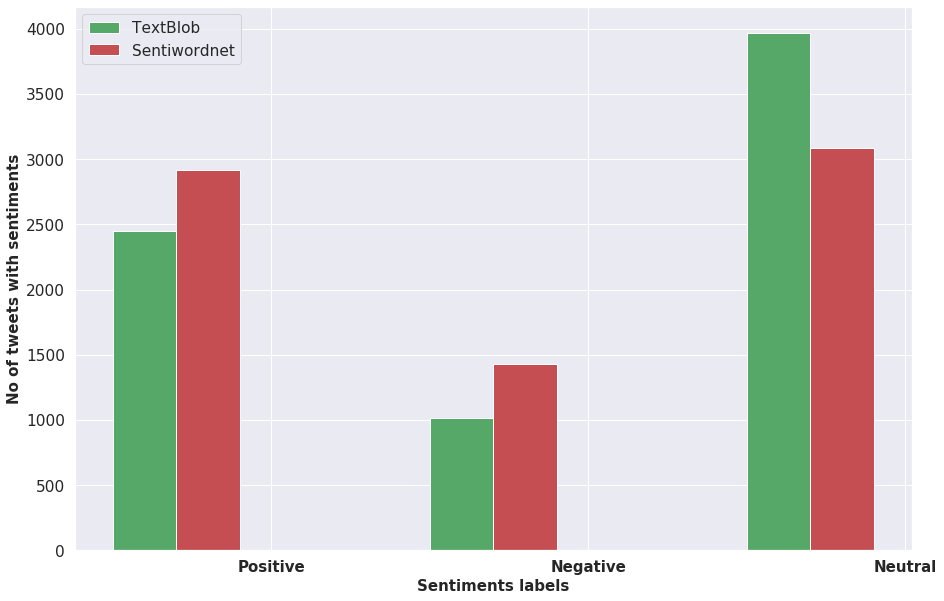}
\caption{\scriptsize Polarity calculation with each sentiment classifier.}
\label{fig:classifiers_c}
\end{figure}

Tweets gathered from public accounts were 33,502 in number. However, after pre-processing only 7430 tweets remained. Among the two sentiment analyzers we compared in this research, we found that SentiWordNet had the highest rate of tweets with positive sentiment, 2916 in number and 39.25\% in percentage. While Textblob is highest in neutral sentiment rate of 3971,  53.45\%, which can be viewed in Table \ref{tab:sentiscores} and Figure \ref{fig:classifiers_c}.

\begin{table}[!htb]
\caption{\scriptsize Percentage/number of polarity calculations of different sentiment classifiers.}
\label{tab:sentiscores}
\centering
\begin{tabular}{l|c|c|c}
\hline
Sentiment Classifier	&	Positive 				& 	Neutral  						& Negative 		\\
\hline
TextBlob				&	2447 (32.93\%)			&	3971 (53.44\%)					& 1012 (13.62\%) \\
SentiWordNet			&	2916 (39.25\%) 			&	3085 (41.52\%)					& 1429 (19.23\%) \\
\hline
\end{tabular}
\end{table}

We use Textblob sentiment tools beyond this point because of its popularity. There are two aspects of Polarity Sentiment Analysis (PSA) and Subjectivity Sentiment Analysis (SSA) conducted. Firstly, we apply both PSA and SSA on all the tweets regardless of the time of tweeting from the users. Secondly, we considered time as a useful dimension of sentiment analysis. The second elucidates the research question 1, considering time-series topic tracking and to find whose name is most mentioned in each topic. In both sentiment analyses, we want to find the attitude of the public towards the political actors and possibly the reason(s).

\subsubsection{Polarity Sentiment Analysis (PSA)}
\label{psa}

\textbf{First PSA}: Regardless of the time of tweets by the users, we compute the sentiment polarity for each tweet in Table \ref{tab:dataset} and aggregate the summary statistics per collection. This analysis includes all the political actors mentioned in section \ref{AGEactors} and in Table \ref{tab:dataset}. Using algorithm 1, we compute polarity scores between +1 to -1. A tweet is classified \textit{positive} if $polarity score > 0$ or \textit{negative} if $polarity score < 0$ otherwise classified as \textit{neutral}. To visualize the overall public opinions or feelings about the election, we compute the sentiment frequency distribution on the overall tweets and per category as recorded in our dataset in Table \ref{tab:dataset}.

\begin{figure}[!htb]
\centering
\includegraphics[width=0.5\textwidth]{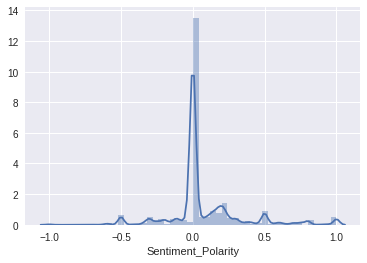}
\caption{\scriptsize The distribution of polarity in our Twitter dataset}
\label{fig:sentianalysis1}
\end{figure}

Figure \ref{fig:sentianalysis1} shows the frequency distribution of sentiment polarity in our dataset. From this Figure, it is evident that most of the tweets in our dataset are positive and have polarity between 0 and 0.5. 

\begin{figure}[htb]
\begin{subfigure}{0.5\textwidth}
  \centering
  \includegraphics[width=0.8\textwidth]{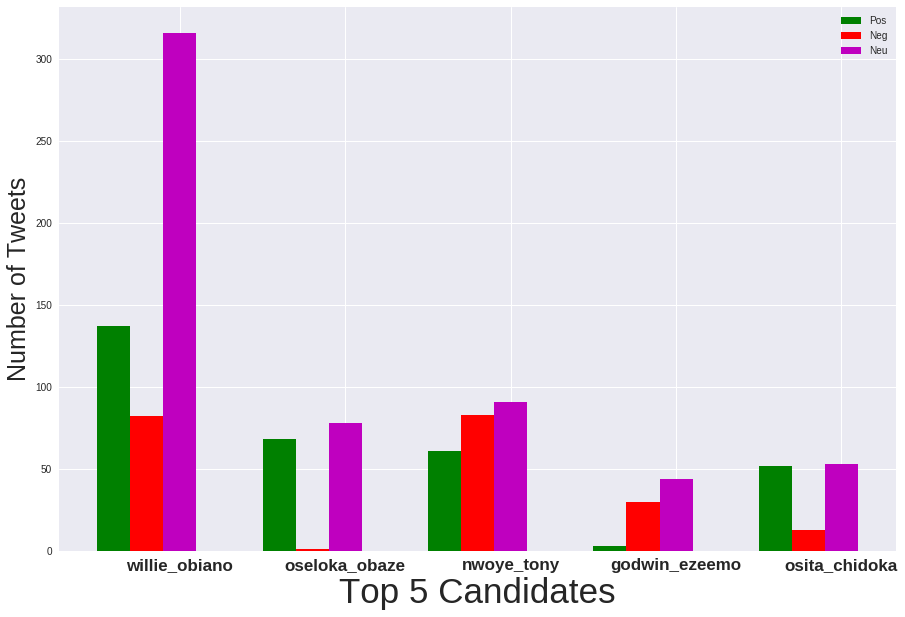}
  \caption{}
  \label{fig:sentianalysis2}
\end{subfigure}
\begin{subfigure}{0.5\textwidth}
  \centering
  \includegraphics[width=0.8\textwidth]{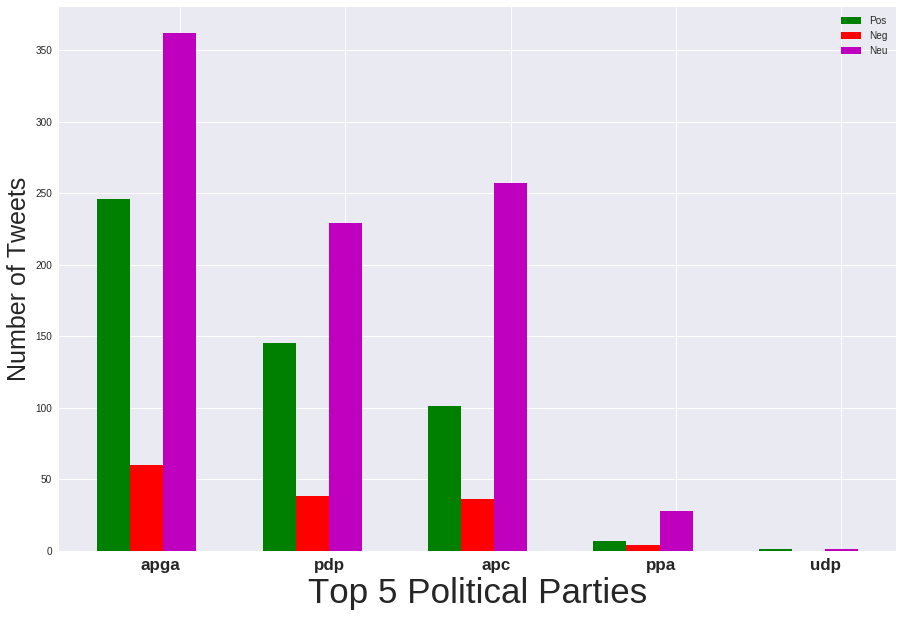}
  \caption{}
  \label{fig:sentianalysis3}
\end{subfigure}
\caption{\scriptsize The polarity sentiment frequency distributions (FD). This shows FD of all the tweets in Table \ref{tab:dataset} where the political actors are mentioned. We used the top 5 political candidates and their parties. (a) is the polarity sentiment frequency distribution per candidate and (b) is the the polarity sentiment frequency distribution per political party}
\label{fig:sentilysis}
\end{figure}

Figures \ref{fig:sentianalysis2} and \ref{fig:sentianalysis3} show the polarity sentiment frequency distribution for the political actors in our dataset categories of Table \ref{tab:dataset}. Figures \ref{fig:namecounts}, \ref{fig:sentianalysis2} and \ref{fig:sentianalysis3} show the political actors in the Anambra State gubernatorial election conducted on November 18, 2017, and their various scores in frequency and sentiment polarity. The frequency distributions of the tweets in these experiments considered tweets where the political actors are mentioned regardless if there are more than one actor mentioned in the same tweet. For example, Figure \ref{fig:sentianalysis3} is a count of tweets that their polarity has been identified. The counts include where the names of the political actors are mentioned irrespective of how many of them appear in a tweet. For example, this tweet ``\textit{Anambra Poll: Election observers, APGA, UPP commend timely distribution of materials}''\footnote{referring to election materials.} from our dataset is positive polarity and is counted for both APGA and UPP political actors. 

\noindent \\
\textbf{Second PSA}: Time is considered as a useful dimension of sentiment analysis. To answer \textit{research question 1}, we used our Twitter dataset grouped according to time of tweet to perform polarity analysis of tweets mentioning each of the political actor. In this phase, we select the top three of the candidates and their parties mentioned in section \ref{AGEactors} to constitute our political actors set. This set is \textit{Willie Obiano (the incumbent Governor) of the state ruling All Progressive Grand Alliance (APGA), Tony Nwoye of the national ruling All Progressives Congress (APC)} and \textit{Oseloka Obaze of the People's Democratic Party (PDP)}. The reason for this selection is considered by the number of tweets mentioning political actors and popularity. For each polarized tweet computed using algorithm 1, we find the time it was tweeted, whose name is mentioned ``solely'' in the tweet at the time and finally compute the average polarity scores of the collection against the political actor's name mentioned. This is to track how people's attitude towards a political actor changes overtime during the election. Time arrangement is based on two hourly granularity starting from 06:00 to 23:59 on the election day. 

\begin{figure}[!htb]
\centering
\includegraphics[width=0.7\textwidth]{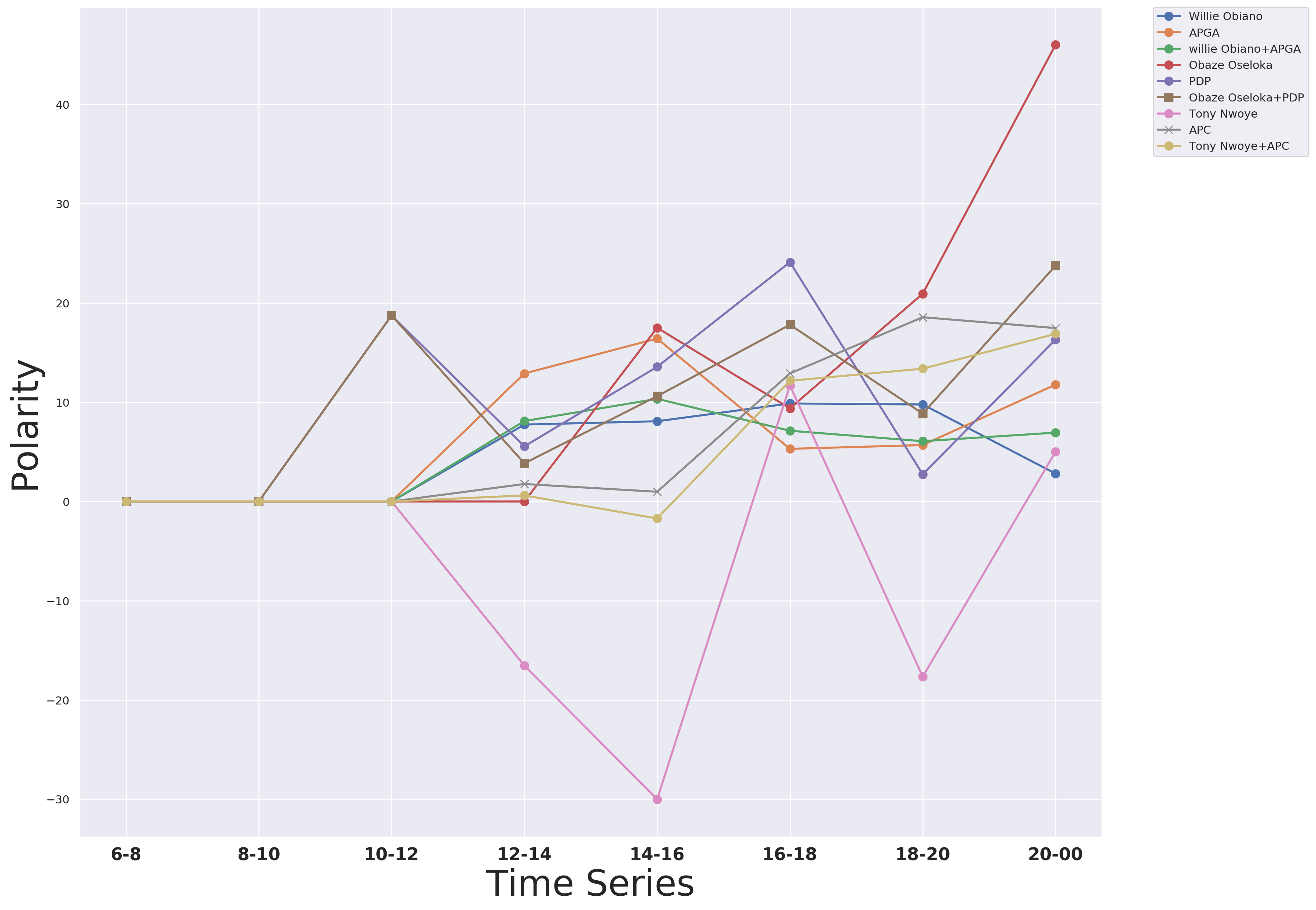}
\caption{ \scriptsize Polarity scores for each political actors based on time scaled by a factor of 100. Time is ranged in two hourly granularity. This graph is used to reveal the attitude of the people towards the political actors in a given time}
\label{fig:polaritySc}
\end{figure}

\noindent Figure \ref{fig:polaritySc} is a graph of the polarity scores on each of the political actor scaled by a factor of 100. It reveals what people are feeling about the political actors in a given time window, what topics are being discussed in each of the time and whose name is mentioned in those topics. Here we can observe interesting patterns such as between \textit{6-8 -- 8-10} there was no tweet specifically mentioning the political actors. But there are general tweets such as 

\begin{quote}
\textit{Ndi Anambra, the next 4 years is critical. It will either be more development for the State or statue building leaders. Please vote wisely. \#AnambraDecides2017}
\end{quote}

\begin{figure}[!htb]
\centering
\begin{subfigure}{0.8\textwidth}
  \centering
  \includegraphics[width=0.9\textwidth]{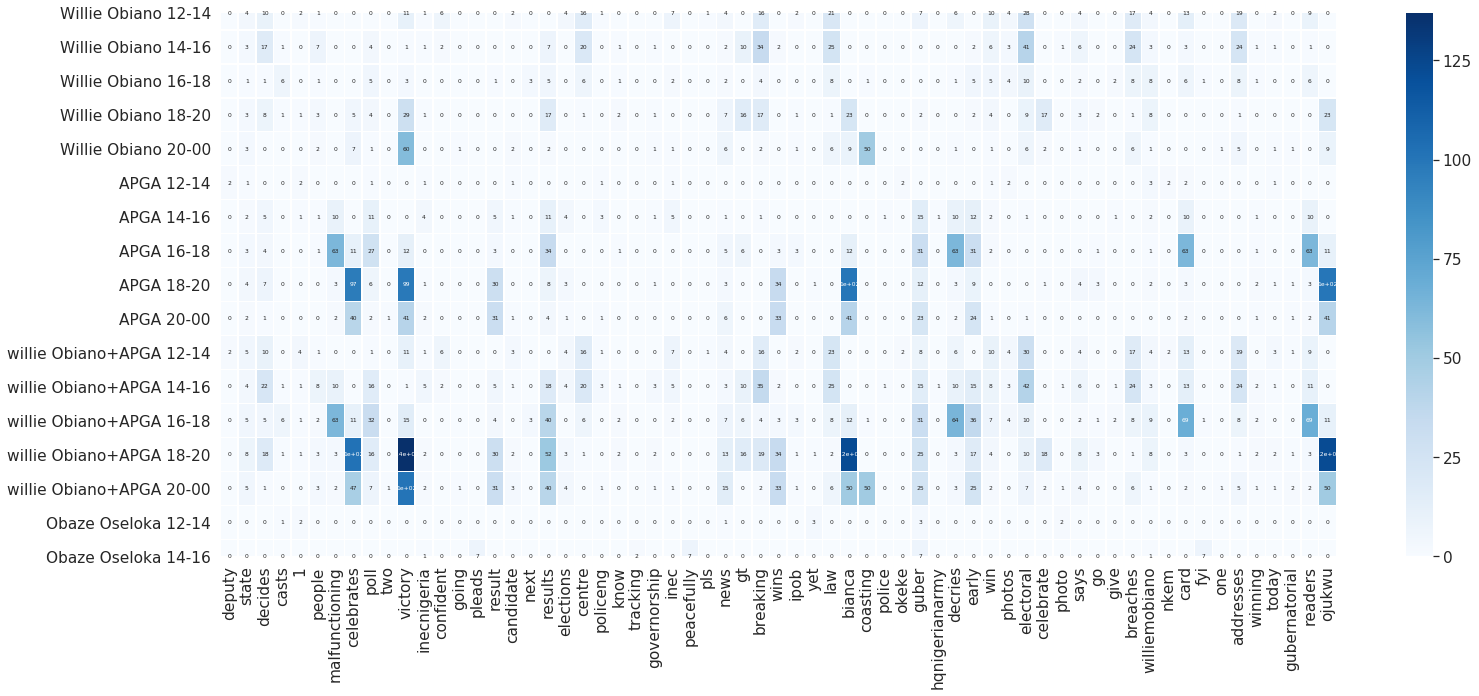}
  \caption{}
  \label{fig:wotopics1}
\end{subfigure}
\begin{subfigure}{0.8\textwidth}
  \centering
  \includegraphics[width=0.9\textwidth]{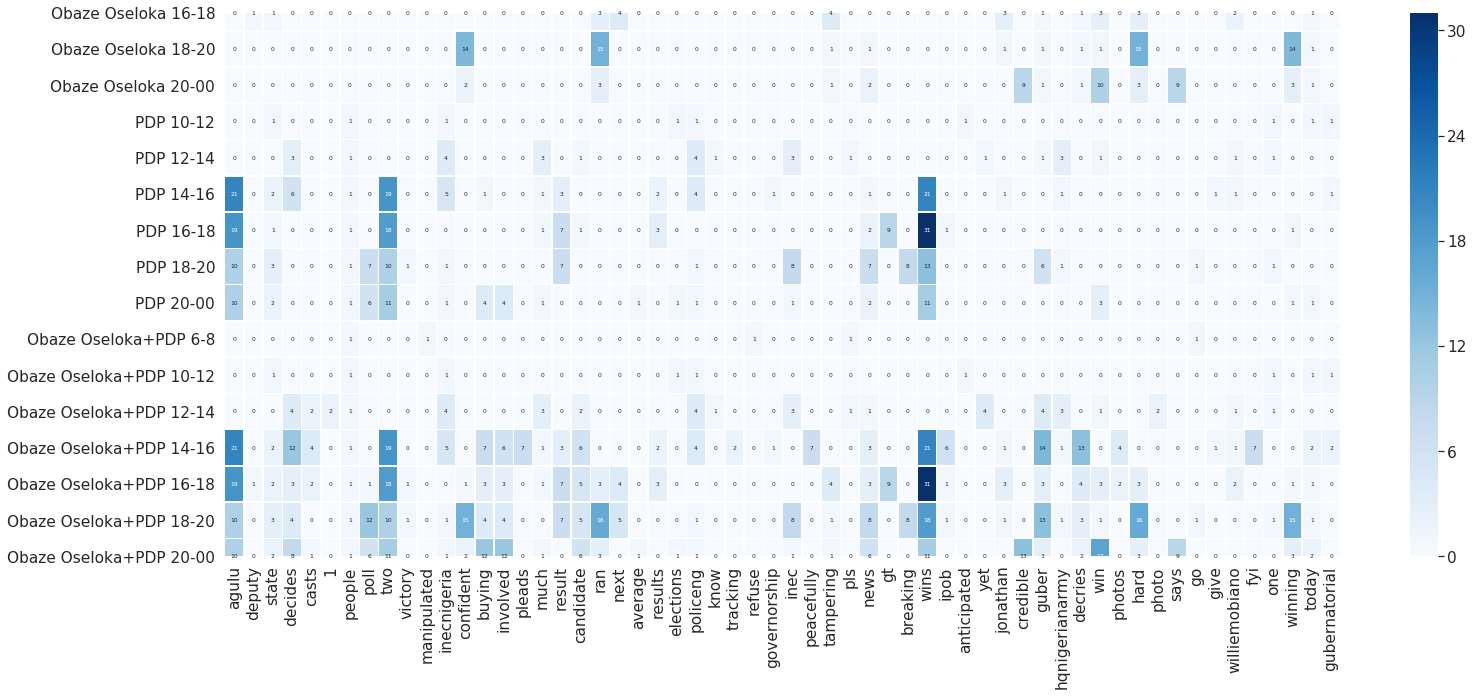}
  \caption{}
  \label{fig:wotopics2}
\end{subfigure}
\begin{subfigure}{0.8\textwidth}
  \centering
  \includegraphics[width=0.9\textwidth]{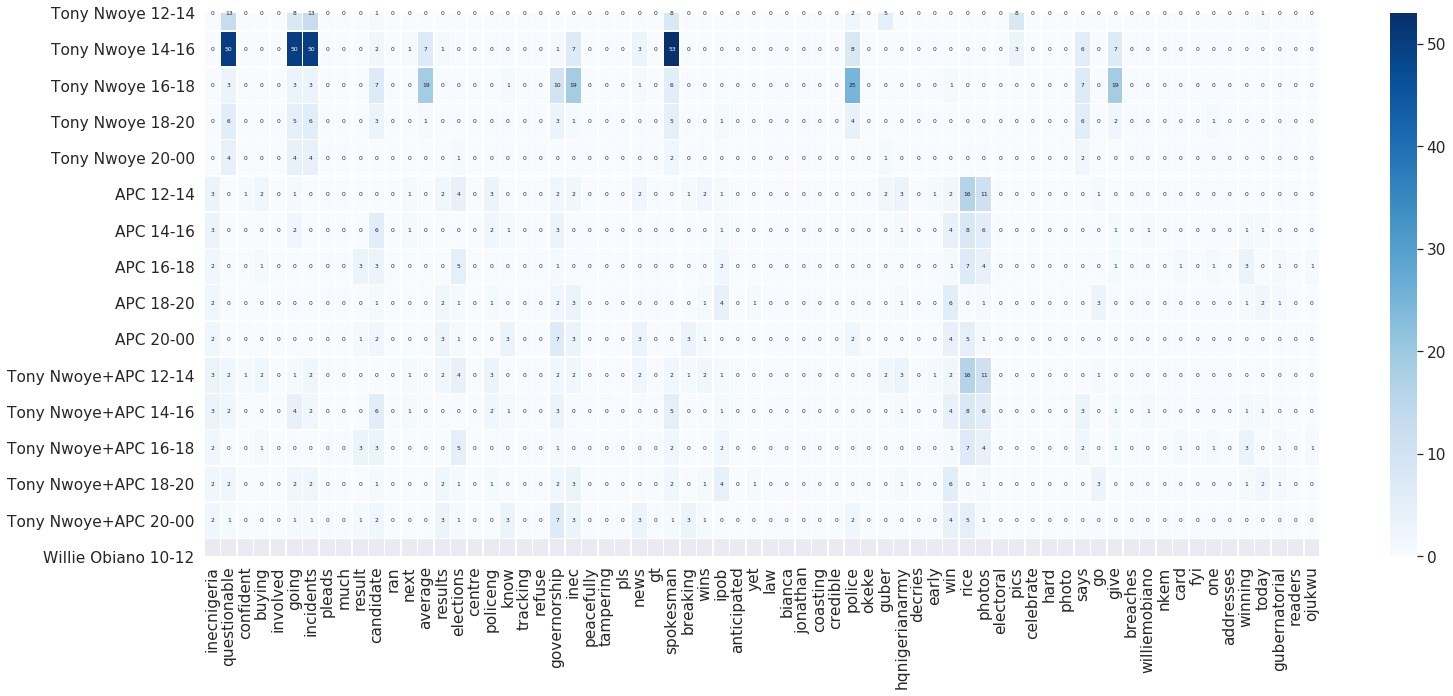}
  \caption{}
  \label{fig:wotopics3}
\end{subfigure}

\caption{\scriptsize Most frequent words based on different time of the election. (a) is for \textit{Willie Obiano},\textit{ APGA}, \textit{Obiano combined with APGA}. (b) is for \textit{ Oseloka Obaze}, \textit{PDP}, \textit{Obaze combined with PDP}. (c) is for \textit{Tony Nwoye}, \textit{APC} and \textit{Nwoye combined with APC}. Time is categorized into two hourly granularity starting from \textit{06:00 to 23:59}. This heatmap is used to discover insights that reveal different most frequent words on each of the political actors across different time during the election fit into topics in Figure \ref{fig:ldatopics}. For example, \textit{Willie Obiano 12-14} showing is the most frequent keywords associated with \textit{Willie Obiano} between 12 pm to 2 pm.} 
\label{fig:wotopics}
\end{figure}

\begin{figure}[!htb]
\centering
\begin{subfigure}{0.9\textwidth}
  \centering
  \includegraphics[width=0.9\textwidth]{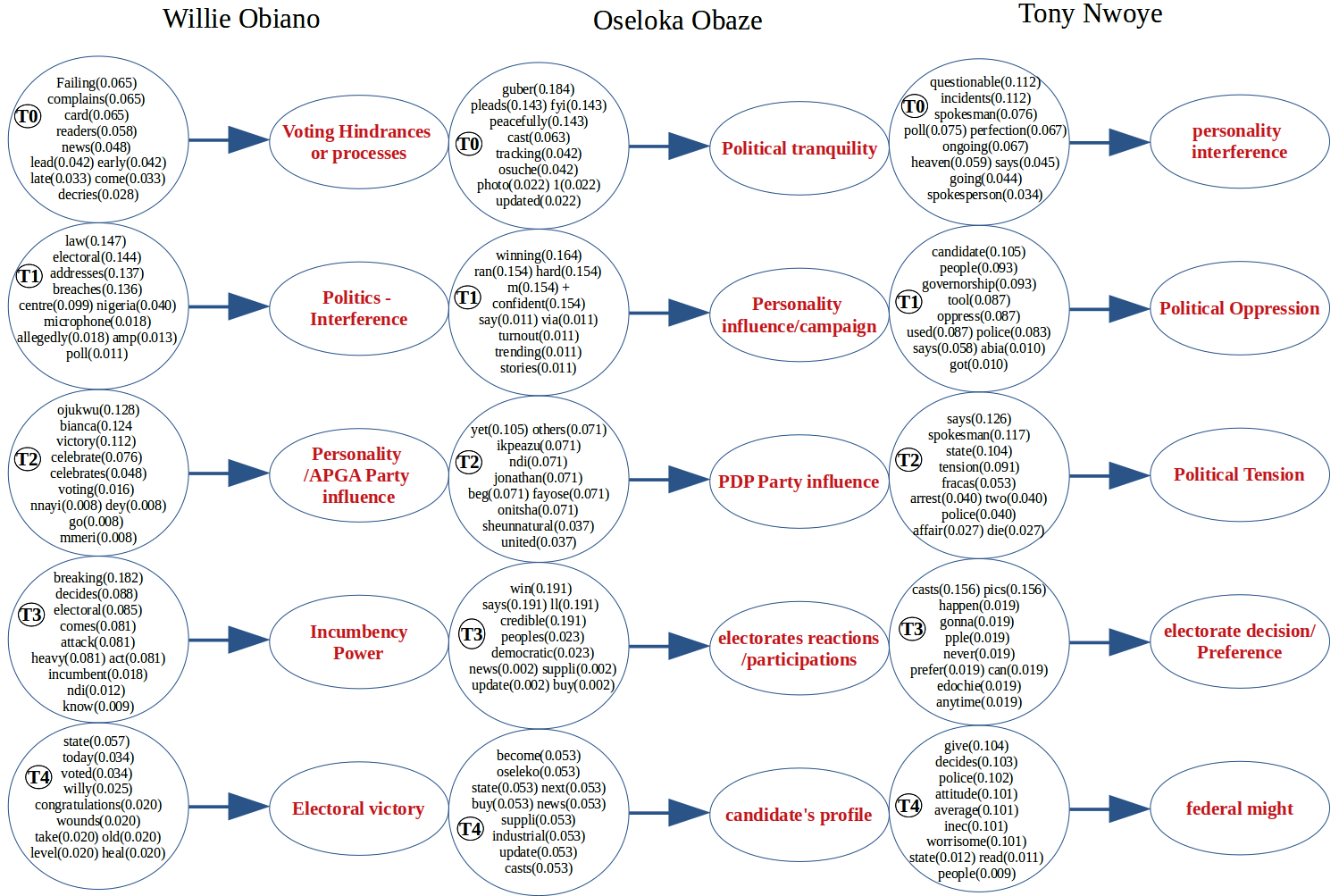}
  \caption{}
  \label{fig:ldatopics1}
\end{subfigure}
\begin{subfigure}{0.9\textwidth}
  \centering
  \includegraphics[width=0.9\textwidth]{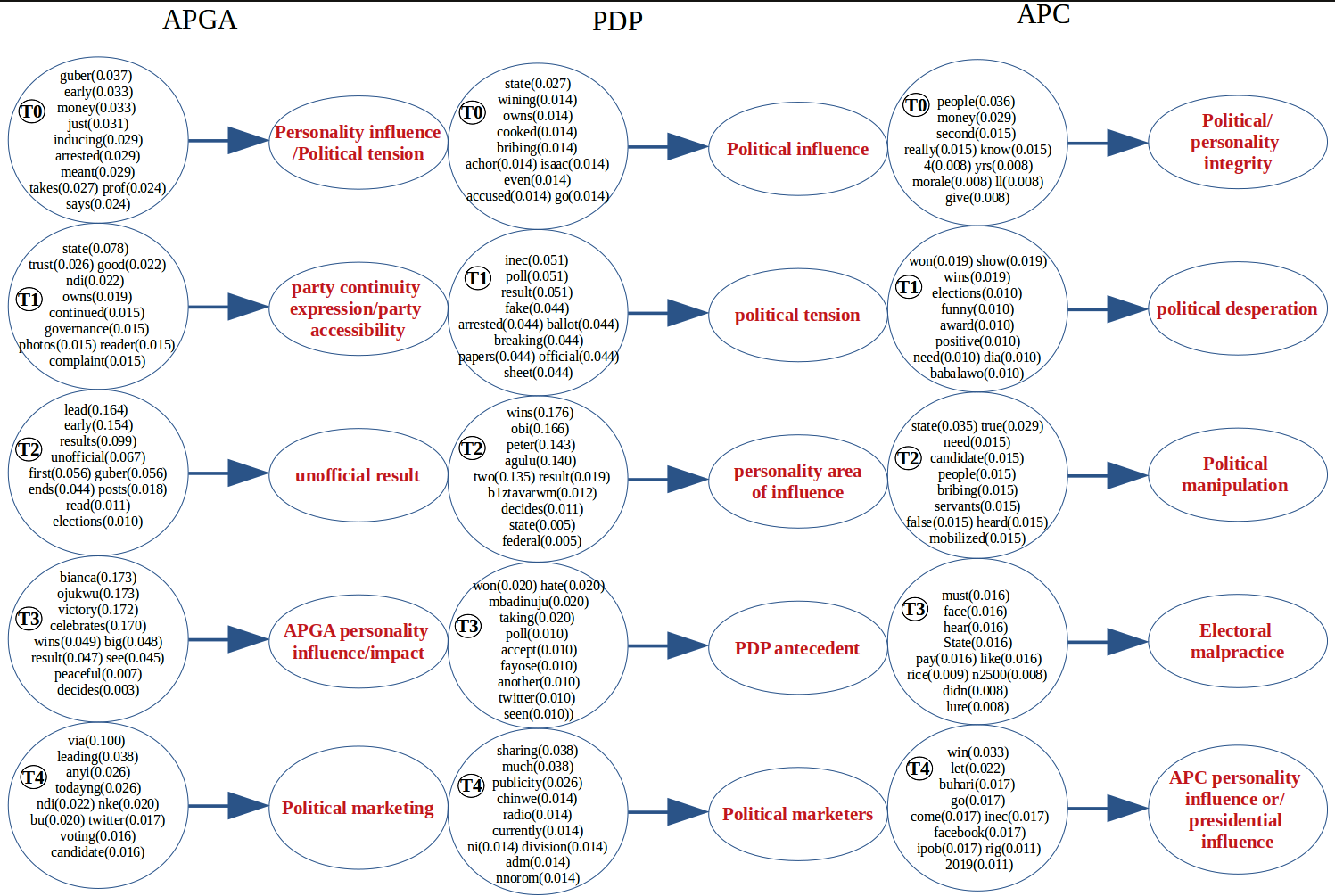}
  \caption{}
  \label{fig:ldatopics2}
\end{subfigure}
\caption{ \scriptsize The top 5 keywords computed by LDA model from tweets where political actors are mentioned and the inferring topics from the keywords. The number of words in easch keyword is 10. (a) is for the political candidates. (b) is for the political parties}
\label{fig:ldatopics}
\end{figure}

\noindent The data points on Figure \ref{fig:polaritySc}, from \textit{8-10 -- 20-00}, are the sentiment polarity scores of the tweets mentioning the political actors. They are often inferred as positive, neutral or negative and from the sign of the polarity score, a tweet is defined as either positive or negative feedback. Thus, they could be used to show the attitude of the public towards the political actors. Generally, from the figure, the tweets mentioning the following political actors \{\textit{Willie Obiano, APGA, Obaze Oseloka, PDP}\} are above the zero bar all through the time from \textit{6-8 -- 20-00}. While \{\textit{Tony Nwoye, APC}\} political actors are below the zero bar, signifying negative comments, at \textit{14-16} for \{\textit{APC, Tony Nwoye}\} and at \textit{12-14} and \textit{18-20} for \textit{Tony Nwoye}. Compare \textit{Tony Nwoye 14-16} on Figure \ref{fig:wotopics} and \textit{Tony Nwoye} at \textit{14-16} data point on Figure \ref{fig:polaritySc}. 

Figure \ref{fig:wotopics} shows the most frequent words associated with the political actors in a given time. It gives insight to what people are saying about them at that time. Figure \ref{fig:ldatopics} shows how important those most frequent words in Figure \ref{fig:wotopics} are in a given topic. Here, topics are computed by a topic model called LDA. It shows the keywords for each topic and the weightage (importance) of each keyword.

Figure \ref{fig:ldatopics} comprises top 5 different topics per political actor built with an LDA model where each topic is a combination of keywords and each keyword contributes a certain weightage to the topic. Adopting the methodology we explained in section \ref{exptools}, we built the LDA model from the corpus and dictionary generated from grouping our tweet data collections into the following political actors, viz; \textit{Willie Obiano (the incumbent Governor) of the state ruling All Progressive Grand Alliance (APGA), Tony Nwoye of the national ruling All Progressives Congress (APC)} and \textit{Oseloka Obaze of the People's Democratic Party (PDP)}. This grouping is based on tweets where these political actors are mentioned. Looking at the \textbf{T0 (topic 0)} under \textbf{Willie Obiano} on the Figure, it means the top 10 keywords that contribute to the topic ``Voting Hindrances'' are: `failing', `complains', `card','reader', ... and the weight of `failing' on \textbf{T0} is 0.065. The weights reflect how important a keyword is to that topic. Looking at these keywords, we guessed what the topic could be by summarising it as ``Voting Hindrances'' associated to the political actor \textbf{Willie Obiano}.

Comparing Figures \ref{fig:wotopics} and \ref{fig:ldatopics}, the x-axis and y-axis makers of Figure \ref{fig:wotopics} show words like \textit{ojukwu} and \textit{bianca} that occurred at \textit{Willie Obiano} and \textit{APGA} with that of  \textit{APGA} showing a high frequency of occurrence than \textit{Willie Obiano}. Also, in Figure \ref{fig:ldatopics}, these two words are very important, considering their weights, in the formation of topics \textbf{T2} and \textbf{T3} of \textit{Willie Obiano} and \textit{APGA} respectively.  \textit{Ojukwu} was a hero in Igbo land and from Anambra State. He was \textit{bianca}'s husband and the founder of APGA. The same could be said of \textit{oseloka obaze} and \textit{PDP} where words such as \textit{agulu, jonathan, obi} in Figure \ref{fig:wotopics} constitute to the formation of topics in Figure \ref{fig:ldatopics} (see \textbf{T2} under both \textit{oseloka obaze} and \textit{PDP}). The high-frequency occurrence of the word \textit{agulu} (also see Figure \ref{fig:wordcl_obazepdp}), just like \textit{ojukwu}, reveals the connection between the candidate and one of the PDP's `kingpin' who hails from Agulu, a town within Anambra state.

Comparing Figures \ref{fig:polaritySc} and \ref{fig:wotopics}, the sentiment polarity of \textit{oseloka obaze} reveals more positive sentiments than the other political actors between \textit{18-20} and \textit{20-00}. Words like \textit{confident}, \textit{ran}, \textit{hard}, \textit{winning}, \textit{credible}, \textit{news}, \textit{win}, \textit{says}, etc., can be observed at \textit{18-20} and \textit{20-00} from Figure \ref{fig:wotopics} showing people's thoughts concerning \textit{Oseloka Obaze}. Also, from Figure \ref{fig:wotopics} at \textit{14-16} and \textit{16-18}, we observed words such as \textit{questionable}, \textit{going}, \textit{incidents}, \textit{spokesman}, \textit{police}, \textit{candidate}, \textit{average}, \textit{governorship}, \textit{inec}, etc., associated with \textit{Tony Nwoye}. And words such as \textit{decides}, \textit{law}, \textit{electoral}, \textit{breaches}, \textit{addresses}, \textit{victory}, \textit{bianca}, \textit{coasting}, \textit{celebrates}, \textit{ojukwu} etc., are found associated with \textit{Willie Obiano} at \textit{16-18}, \textit{18-20} and \textit{20-00} on Figure \ref{fig:wotopics}. Most of these words formed top 10 important words used in the formation of topics in Figure \ref{fig:ldatopics} for the political actors \textit{Willie Obiano, Tony Nwoye, Oseloka Obaze, APGA, APC, PDP}. 

\subsubsection{Subjective Sentiment Analysis (SSA)}

An objective sentence expresses some factual information about the world, while a subjective sentence expresses some personal feelings or beliefs.  For  example,  the sentence ``\textit{This past Saturday, I bought a Nokia phone and my girlfriend bought a Motorola phone}'' does not express any opinion hence is objective, while  ``\textit{The  voice  on  my  phone  was  not  so  clear,  worse  than  my  previous  phone}'' sentence is subjective.  Subjective expressions come in many forms, for example, opinions, allegations, desires, beliefs, suspicions, and speculations \cite{Riloffetal2006,Wiebe2000}. Thus, a subjective sentence may not contain an opinion. For example, ``\textit{I wanted a phone with good voice quality}'' is subjective but it does not express a positive or negative opinion on any specific phone. Similarly, we should also note that not every objective sentence contains no opinion as in ``\textit{The voice quality of this phone is amazing}'' \cite{liu2010sentiment}. The issue of subjectivity has been extensively studied in the literature \cite{Hatzivassiloglou1997,Hatzivassiloglou2000,Riloffetal2006,Wiebe2000,Riloffetal2003,liu2010sentiment}.

\begin{figure}[htb]
\begin{subfigure}{0.5\textwidth}
  \centering
  \includegraphics[width=1\textwidth]{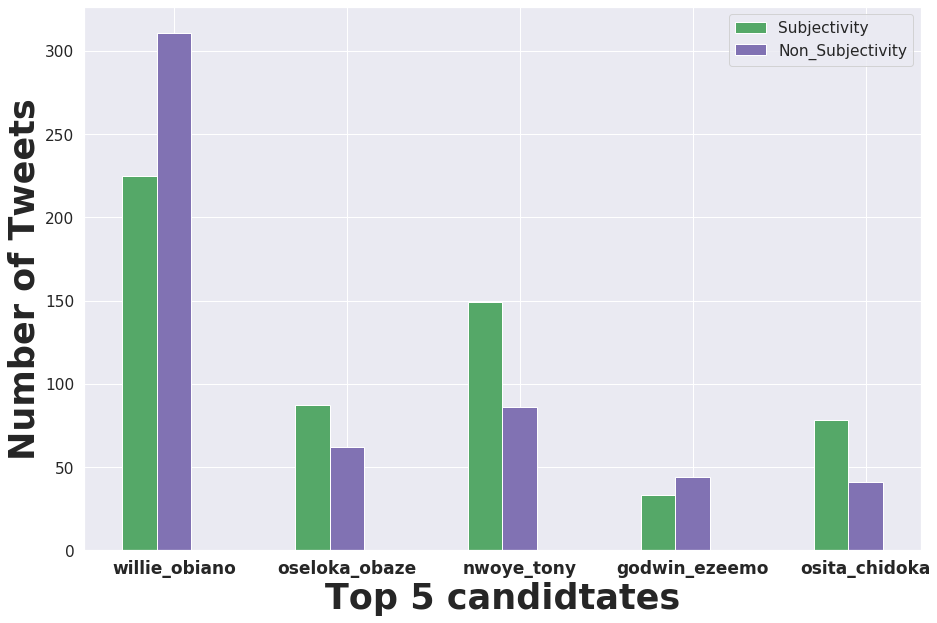}
  \caption{}
  \label{fig:sentianalysis2subj}
\end{subfigure}
\begin{subfigure}{0.5\textwidth}
  \centering
  \includegraphics[width=1\textwidth]{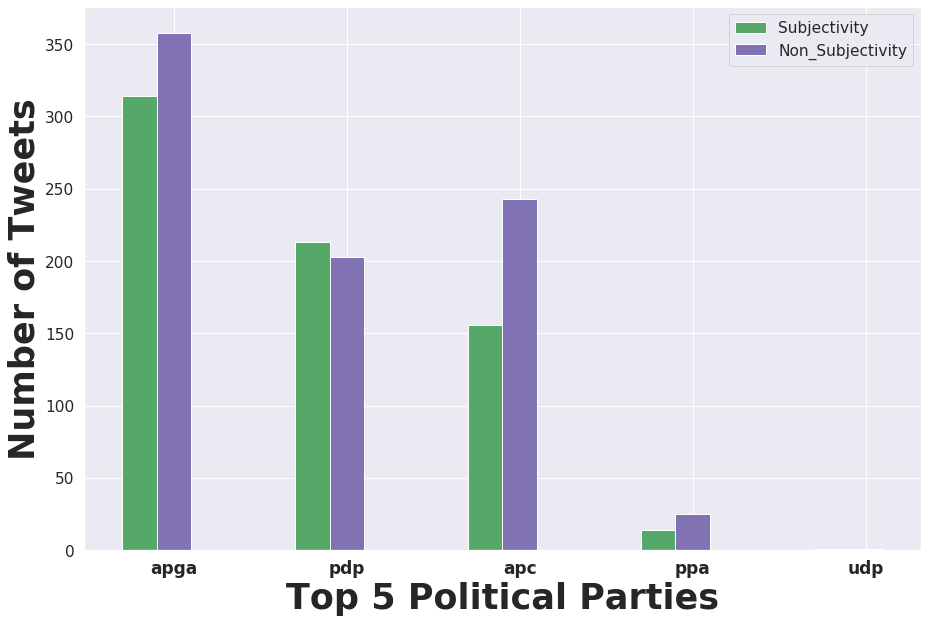}
  \caption{}
  \label{fig:sentianalysis3subj}
\end{subfigure}
\caption{\scriptsize The subjectivity sentiment frequency distributions (FD). This shows FD of all the tweets in Table \ref{tab:dataset} where candidates, political parties or both are mentioned. We used top 5 political candidates and parties. (a)-The subjectivity sentiment frequency distribution per candidate. (b)-The subjectivity frequency distribution per political party}
\label{fig:sentilysissubj}
\end{figure}

\begin{figure}[!htb]
\centering
\includegraphics[width=0.7\textwidth]{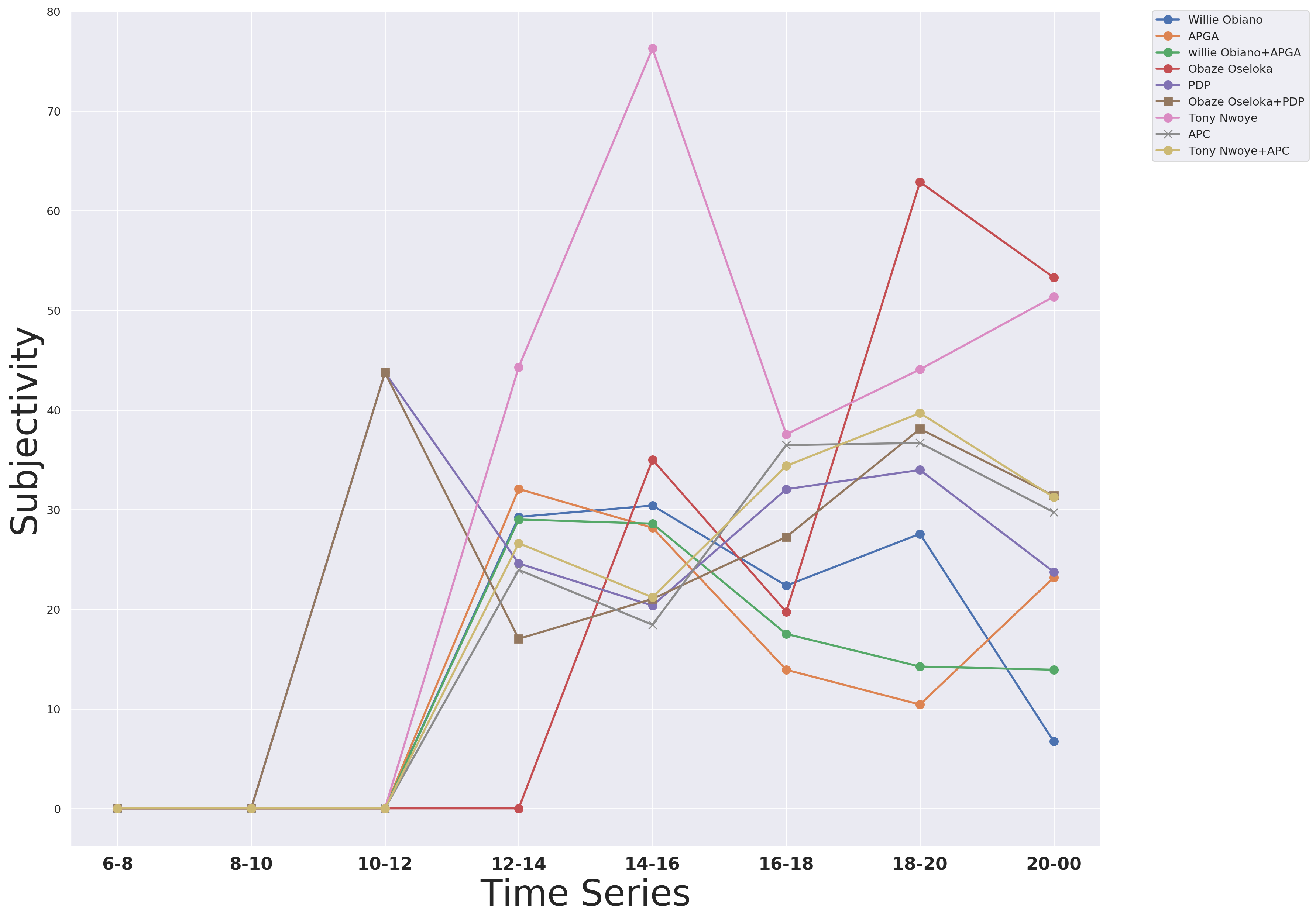}
\caption{\scriptsize Subjectivity scores for each political actors based on time. Time is formatted in two hourly granularity from \textit{06:00 to 23:59}}
\label{fig:subjectivitySc}
\end{figure}

\textbf{First SSA}: We perform SSA on our Twitter data in Table \ref{tab:dataset} without time consideration. This is to evaluate the overall Twitter texts tweeted during the Anambra State gubernatorial election conducted on November 18, 2017, whether they are factual or emotional subjective opinions. The evaluation involves all the candidates and their political parties mentioned in section \ref{AGEactors}. While sentiment polarity in section \ref{psa} determines the positive or negative connotation of a tweet in our Twitter dataset, SSA tries to discern whether the tweet is subjective in the form of an opinion, belief, emotion or speculation or objective as a fact. Thus, we investigate tweets mentioning the political candidates, parties or both to know their subjectivity scores and which one of them is higher. This is illustrated in Figure \ref{fig:sentilysissubj}. We can observe that tweets mentioning \textit{willie\_obiano} and \textit{apga} are more non--subjective than others. A similar case can be seen in \textit{godwin\_ezeemo} and \textit{ppa} except in \textit{godwin\_ezeemo\_ppa}, tweets mentioning both names,  where both subjectivity and non--subjectivity are equal. \textit{Oseloka Obaze} and his party \textit{PDP} tweets are more subjective as revealed in all the bar plots. A dissimilar trend is observed in \textit{tony\_nwoye} and his party, \textit{APC} where the tweets mentioning the candidate's name is more subjective, the tweets mentioning  the party is more non-subjective and the tweets mentioning both names are more non--subjective. The non-subjectivity of the latter can be viewed as an influence from the party's non--subjectivity results.

\noindent \\
\textbf{Second SSA}: In this analysis, we considered time as a useful dimension of sentiment analysis. Along with polarity analysis \textit{PSA} above, we have also used the same data and time arrangement to answer  \textit{research question 2}. The results for the SSA are shown in Figure \ref{fig:subjectivitySc}. 

We observed from the figure that tweets `solely' mentioning \textit{Tony Nwoye} are more subjective between \textit{14-16} time. A similar case is observed of \textit{Oseloka Obaze} but between the \textit{18-20} time group. \textit{Willie Obiano} and his party, \textit{APGA} started on subjective scores a bit lower than others but at the end of the day, they got the lowest scores than others showing that tweets mentioning their names are less opinionated. Again, we can say that their results are closely knitted at morning time between \textit{10-12} and at afternoon time between \textit{14-16 -- 16-18}.  Furthermore, it could be envisaged that the personalities of the candidates \textit{Oseloka Obaze} and \textit{Tony Nwoye} drove people to be emotionally subjective in their tweets about them leading to the huge differences between their subjectivity scores and that of their parties.

A high subjectivity score does not indicate a higher propensity for a voter to vote. What it does indicate however is that Twitter users mentioning \textit{Tony Nwoye} and \textit{Oseloka Obaze} in their tweets are more opinionated, and those mentioning \textit{Willie Obiano} are less opinionated.

\section{Discussion}
From the above figures and analyses, we deduce that even though a political party serves as a platform that sales the personality of a political actor or contestant while struggling for power, the credibility of a political actor may even though add strength to the spread of the party. For example, Figures \ref{fig:sentilysis},  \ref{fig:polaritySc} and \ref{fig:wotopics} reveal attitudes of the public towards the political actors. The political actors \textit{Oseloka Obaze} and \textit{Osita Chidioka} are well accepted than their political parties in either positive polarity or popularity. The variable \textit{oseloka\_obaze} on Figure \ref{fig:sentianalysis2} shows a negative polarity score that is almost zero (also see \textit{osita\_chidioka}). Most frequent words such as \textit{confident},  \textit{credible}, \textit{news}, \textit{win}, \textit{says}, \textit{peacefully},  etc., can be observed from Figure \ref{fig:wotopics2} showing people's thoughts concerning \textit{Oseloka Obaze}. However, the viability and acceptability of a given political party to the electorates have a greater effect on the victory of the party and the candidate it presents. Individual efforts by the political actors in promoting their political manifestoes through the political party that does not win the sympathy of the electorates are usually of less effect in actualizing political victory especially in a developing nation like Nigeria. This is illustrated in the case of \textit{Willie Obiano} and his party \textit{All Progressive Grand Alliance (APGA)} as explained in the paragraphs below.

Political behaviour of the electorates during an election has a connection to the political party that has an ideological link to their belief, culture and value. For instance, in Nigeria PDP, could rule Nigeria for 16 years (1999-2015) was not because of the power of incumbency rather also as a result of its acceptability to the people notwithstanding who is her flagbearer. Muhammed Buhari contested presidential elections in 2003, 2007 (All Nigerian People's Party-ANPP), 2011 (Congress for Progressive Change- CPC) but lost majorly because of the poor spread of his party as a result of unacceptability to the electorates not minding his personality. This resembles the case of the political actor \textit{Godwin Ezeemo} and political parties \textit{PPA} and \textit{UPP} as shown in their sentiment scores and tweets frequencies in Figures \ref{fig:sentianalysis2}, \ref{fig:sentianalysis3} and \ref{fig:namecounts}. In 2013, Buhari formed an alliance with other political parties (Action Congress of Nigeria- ACN; Congress for Progressive Change- CPC; All Nigerian People's Party- ANPP; a faction of All Progressive Grand Alliance- APGA and aggrieved members of Peoples Democratic Party- PDP). This alliance gave rise All Progressive Congress (APC) with wider coverage and acceptability in the North-East, North-West, North-Central and South-West of Nigeria. In view of this, Buhari that lost election three consecutive times (2003, 2007 and 2011) won the 2015 general election against the incumbent president (Goodluck Jonathan of PDP). 

\begin{figure}[!htb]
\centering
\includegraphics[width=0.8\textwidth]{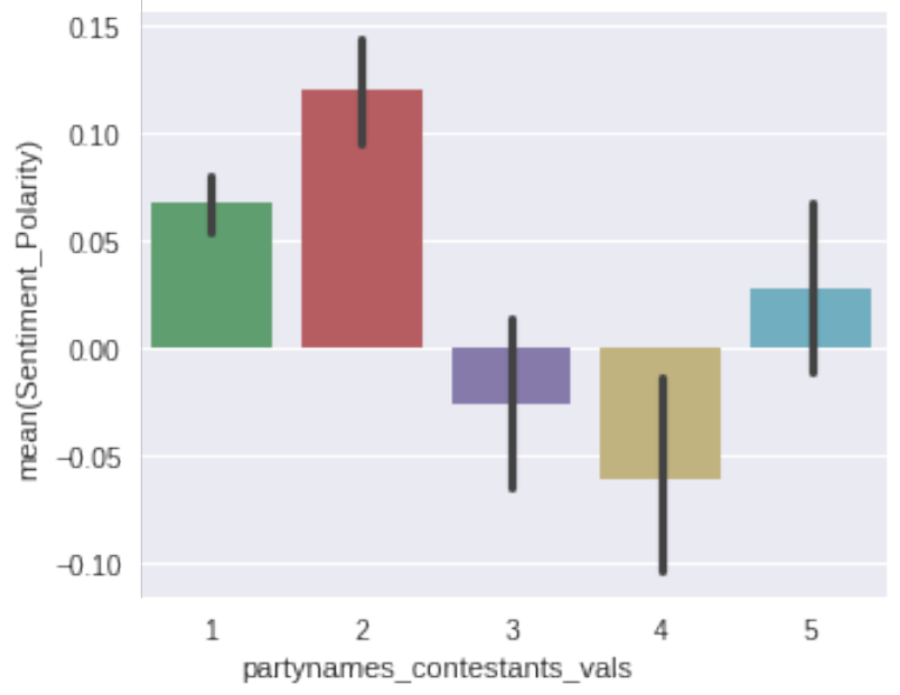}
\caption{The average sentiment polarity for each political party cum candidate. Numbers 1,2,3,4,5 represents APGA, PDP, APC, PPA and UPP respectively}
\label{fig:sentianalysis5}
\end{figure}

Empirical observations on the political actor \textit{Willie Obiano} and his party \textit{APGA} from Figures \ref{fig:sentianalysis2}, \ref{fig:sentianalysis3} and \ref{fig:wordfreqCloud} reveal that the political-ideological links such as people's belief, culture, value and acceptability add to the \textit{Willie Obiano} winning the November 18, 2017, Anambra State gubernatorial election. From Figure \ref{fig:wordfreqCloud}, the most frequent outstanding words associated with \textit{Willie Obiano} when combined with \textit{APGA} can be connected to the ideology of the Igbo nation and Chukwuemeka Odimegwu Ojukwu signifying their normal slogan \textit{nke a b\d{u} nke any\d{i}} `this is our own'. Moreso, Figure \ref{fig:ldatopics2}, topic \textbf{T4} (political marketing) can further be explained, based on the topic keywords, to be \textit{indigenous acceptance ``Nke a b\d{u} any\d{i}'' (This is our own)}. Also, from Figures \ref{fig:sentianalysis2} and \ref{fig:sentianalysis3}, the sentiment polarity frequency distribution associated with \textit{Willie Obiano} shows a more negative sentiment compared to his party \textit{APGA}, while Figure \ref{fig:sentianalysis5} shows all positive on the average sentiment polarity scores between \textit{Willie Obiano} and \textit{APGA}. The variable \textit{willie Obiano+APGA} at \textit{18-20} and \textit{20-00} on Figure \ref{fig:wotopics1} diplays more positive words such as \textit{celebrates, victory, results, bianca, coasting, guber, early, ojukwu}, etc., than when only \textit{willie Obiano} is used.  APGA since 2008 till date has been winning gubernatorial elections of Anambra State-Nigeria due to the acceptability of the party to the people and belief on the party as Igbo party. APGA survived the influence of National incumbent parties (PDP and APC) in 2010, 2014 and 2018 gubernatorial elections not because of the qualities of the candidates rather because of the party's influence on the people. The founder of the party APGA (Chukwuemeka Odimegwu Ojukwu) is a generally accepted personality among the Igbo nation, especially his state Anambra (see Figure \ref{fig:wordfreqCloud} for the most frequent words with \textit{Willie Obiano} and \textit{APGA}). The objective of the party as the founder always advocated is to promote Igbo ideology, to unite the Igbo nation within Nigeria state and to have a political umbrella to advance Igbos interest, made the majority of the Anambrarians especially the masses (electorates) to support the party in every election notwithstanding who is its flagbearer. 

In furtherance to our discussion on Figure \ref{fig:subjectivitySc} about objective and subjective opinions, APGA is a ruling political party in Anambra state for 12 years now. It has been able to design policies and execute infrastructure that outwit other political parties like PDP that was in power for 7years (1999-2003), APC, UDP, PPA that have not been in power. This singular opportunity made tweets associated with APGA as a party and its candidate less subjective unlike other political actors. However, other political parties like APC, PDP, PPA UDP and their candidates have not had such opportunity in their political adventure in the state. In view of this, people's connection to these parties are not concrete and there is no concrete policies and projects linked to them in the state. This may be the likely reason why the tweets associated with these parties' candidates are more subjective. See also Figure \ref{fig:sentianalysis2subj} for the subjectivity sentiment frequency distribution on the candidates.

Finally, variable like personality influence occurred repeatedly in Figure \ref{fig:ldatopics}. This shows that tweets of the people are indicating that personality influence plays significant roles in winning an election. Political party with an acceptable personality coupled with its influence as a party does well than a party who has a person of no or little influence in the society or among the people. Therefore, personality influence and political party influence are very important and underlining factors in electoral victory.

\section{Conclusion}
In this research, we investigate how candidates/their political parties could influence winning or losing an election using Twitter data.  We stated research questions that enable us to evaluate our dataset to gain insight on the political phenomenon that could help us in our research.

We tested our research questions by using Twitter data collected during the Anambra gubernatorial election 2017 as a case study. We analyzed over 7k Twitter messages streamed during the election day only. Since Twitter users tweets in real-time, we believe that tweets during the election day are based on what the users are experiencing at that moment and could gain political insight from them. The tweets collected are analyzed exploratively and sentimentally to answer our two research questions. In the explorative experiment, we gained overall insights on our data such as \textit{Willie Obiano} and his party, APGA recording highest number of tweets mentioning their names. This reveals to us the likely reasons behind him/his party having the highest number of positive and negative tweets in the proportion of frequency distributions of the sentiment `polarize and subjective' tweets (see Figures \ref{fig:sentilysis} and \ref{fig:sentilysissubj}). In sentiment analysis, we found people's attitudes towards the political actors across a given set of time and whether these attitudes are subject to facts or opinions. Our tweets collected were segmented into two-hour time groups forming 8 groups starting from \textit{06:00} to \textit{23:59}. The primary purpose of this study was to utilize this time-based information, as a useful dimension for sentiment analysis, contained in the message metadata to create a more detailed analysis of Twitter users subjectivity and polarity during the elections. At this stage, a polarized tweet must contain one of the selected candidates/political parties names in order to be considered. For each time group, we filtered tweets that were tweeted between the set time range and grouped them based on the names they are uniquely mentioning. Then find the average polarity and subjectivity scores of each group within the said time. Generally, the polarity average scores for all selected cases are above bar as shown in Figure \ref{fig:polaritySc} with only \textit{Oseloka Obaze} exceeding 30\% but at 20--00 time, and except \textit{Tony Nwoye} and \textit{APC} are the only cases where the scores are negative. Furthermore, it was observed that tweets mentioning \textit{Tony Nwoye} and \textit{Oseloka Obaze} were more subjective in nature than those mentioning \textit{Willie Obiano} and his party. This is likely because APGA has been a ruling political party in Anambra state for 12 years now. We also use LDA model to build topics in Figure \ref{fig:ldatopics}  to find various topics being discussed and we compared the polarity and subjectivity analyses with the findings. Again, we used Figure \ref{fig:ldatopics}to to check on how important a word in Figure \ref{fig:wotopics} is based on its wieght.

A high subjectivity score does not indicate a higher propensity for a voter to vote. However, it indicates that Twitter users mentioning \textit{Tony Nwoye} and \textit{Oseloka Obaze} in their tweets during the lunch and supper times are more emotional subjective in their messages, whereas those mentioning \textit{Willie Obiano} at the same time are not.

Finally, the Twitter Analysis and Visualization on \#AnambraDecides2017 shows that political actors leverage on the impacts of social media (Twitter, Facebook, WhatsApp, Youtube and other blogs) to define and determine political behaviour of the electorates to win elections. It also adds in validating that political insights are a phenomenon present on social media.


%
\section*{Conflict of interest}
On behalf of all authors, the corresponding author states that there is no conflict of interest. 



\end{document}